\documentclass[twocolumn,showpacs,amsmath,amssymb,
 aps,
 prb,
 floatfix,
 lengthcheck,%
]{revtex4-1}
\usepackage{graphicx}
\usepackage[english]{babel} 
\usepackage[utf8]{inputenc}
\usepackage{hyperref}
\usepackage{color}
\usepackage[normalem]{ulem}

\makeindex
\def\>{\right\rangle}
\def\<{\left\langle}
\def\be{\begin{equation}}
\def\ee{\end{equation}}
\def\ba{\begin{array}{l}}
\def\ea{\end{array}}
\def\beq{\begin{eqnarray}}
\def\eeq{\end{eqnarray}}

\usepackage{xcolor}

\usepackage{cleveref}

\begin{document}

 
\title{Quasiparticle Andreev reflection in the Laughlin fractions\\ of the fractional quantum Hall effect}

\author{K. Iyer}
\email{kishore.iyer@cpt.univ-mrs.fr}
\author{T. Martin}
\author{J. Rech}
\author{T. Jonckheere}
\affiliation{Aix Marseille Univ, Universit\'e de Toulon, CNRS, CPT, Marseille, France}

\date{\today}

\begin{abstract}
Andreev reflection occurs in a normal metal-superconductor junction,
when an electron on the normal side can only be transmitted as a Cooper pair in the superconductor, with the reflection of a hole on the normal side. A similar phenomenon can
occur in strongly correlated systems, in particular in the fractional quantum Hall effect (FQHE), as the system quasiparticles have a charge
$e/m$ different from the electron charge. We study theoretically a setup
involving two quantum point contacts (QPC) in the FQHE where
Andreev reflection occurs, as charges $e/m$ impinging on the second QPC can
only be transmitted as charges $e$, with the reflection of holes of charge
$e (1-m)/m $. Using the bosonization formalism, and out-of-equilibrium
Keldysh Green function techniques, we provide a full analytical calculation of the current correlations at the outputs of the QPC, both at zero and finite temperature. The ratio between the auto- and cross-correlations of the output currents is a direct manifestation of Andreev reflection. Our 
results agree with recent experimental observations, and give precious
information on the temperature dependence of this ratio.
\end{abstract} 


\maketitle

\section{Introduction}\label{sec:intro} 

Andreev reflection~\cite{tinkham96} is usually understood as a paradigm which is proper to superconducting devices. In a normal metal/superconductor junction biased in the subgap regime, while a single electron impinging from the normal metal side cannot be transferred  to the superconductor, a Cooper pair  can be transmitted into the latter provided that a counter-propagating hole is reflected back in the normal metal. This fundamental process of superconducting devices has had huge applications in electronic transport, such as multiple Andreev reflection in two \cite{cuevas96} (and three \cite{Jonckheere2013}) lead superconducting junction, crossed Andreev reflection,\cite{martin96,torres1999,lesovik2001,recher2001,sauret2004,chevallier2011,rech2012} etc.

Strikingly, a phenomenon akin to Andreev reflection  has been also proposed in a totally different context of the strongly correlated state of matter of the Fractional Quantum Hall Effect\cite{tsui99,laughlin83} (FQHE), in quantum Hall bars where two Laughlin filling fractions of the quantum Hall effect coexist.\cite{Sandler1998} This FQHE Andreev reflection was recently demonstrated experimentally in heterostructures where half of the Hall bar has filling fraction $\nu=1$ and the other half has $\nu=1/3$.\cite{Hashisaka2021,Sandler1998, cohen2022} There, the basic signature of Andreev reflection is the enhancement of the conductance when two incoming charges $e/3$ quasiparticles are scattered into a transmitted electron with charge e and a reflected quasi-hole with charge $-e/3$. Such phenomena have also been proposed in the context of interacting one-dimensional wires. \cite{fukuzawa2023minimal}

A different kind of Andreev reflection in the FQHE has been proposed in a single Laughlin quantum Hall bar\cite{Kane2003} with filling fraction $\nu=1/m$ ($m$ odd), endowed with two quantum point contacts (QPCs) as depicted in Fig.~\ref{fig:setup}. The first QPC is tuned to be transparent, and as a result, the back-scattering current consists of a dilute beam of $e/m$ quasiparticles, which then propagates on the opposite edge. On the other hand, the second QPC placed downstream of this dilute beam is opaque, effectively breaking the quantum Hall liquid into two, and hence transmits only electrons. Such a geometry was first studied experimentally by Ref.~\onlinecite{Comforti2002}, who found the charge transmitted across the second QPC to vary as a function of the dilution of the incoming quasiparticle current. Surprisingly, for a highly dilute beam, they found that fractional charges -- contrary  to electrons -- could go across the second QPC, which only electrons are expected to traverse. Later works \cite{Chung2003, Glidic2023} attributed this unexpected result to finite-temperature effects, and possibly to the long propagation path of the quasiparticle beam.

Quasiparticle Andreev reflection in the two QPC geometry in the $\nu = 1/3$ FQHE was recently demonstrated,\cite{Glidic2023} where the ratio of cross- and auto-correlations was measured to be $-2/3$. The experiment managed to achieve this feat by working at very low temperatures and having a short propagation length between the two QPCs. These features ensure that the quasiparticles traveling on the edge do not interact strongly with the surrounding environment, or with each other significantly, before impinging on the second QPC. Their result of a negative cross-correlation noise was distinguished from anyonic braiding effects arising in similar geometries (with both QPCs transmitting $e/3$ quasiparticles)\cite{bartolomei20, Ruelle2022, Glidic2022braiding, Lee2022}  by measuring cross-correlations as a function of the asymmetry of the input current, which remained constant. 

Ref.~\onlinecite{Kane2003} were the first to theoretically address the transmission of dilute Laughlin quasiparticles in the two QPC geometry of Fig.~\ref{fig:setup}. They find that although $e/m$ quasiparticles impinge on the second QPC, the latter transmits electrons. Charge conservation imposes that this process is accompanied by the reflection of $m-1$ quasiholes 
with a total charge  $e/m-e = e (1-m)/m$. Ref.~\onlinecite{Kane2003} uses non-perturbative analytic calculations based on refermionization in the fictitious $\nu = 1/2$ system as well as a mixture of perturbative analytics and numerics for a generic $\nu = 1/m$ system to motivate their results. However, this work is restricted to the evaluation of tunnel observables and lacks a complete analytic evaluation of the auto- and cross-correlation (or even the tunneling noise) for $\nu = 1/m$, whose ratio explicitly demonstrates quasiparticle Andreev reflection.
In the present work, we aim at bridging the gap between theory and experiment -- using the chiral Luttinger liquid theory of the FQHE within the  Keldysh formalism --  by providing a detailed analytic description of the latter phenomenon. 

Analytical approaches which use perturbation theory in tunneling operators of the chiral Luttinger model typically represent a daunting task when dealing with more than one QPC -- see for instance Refs. \onlinecite{safi01,kane03,jonckheere05, Lee19} where this is achieved. Previous works dealing with more than one QPC, such as Fabry-Perot setups,\cite{nayak08} either resort to second order perturbation theory (which is irrelevant for the present problem), or have to use
different techniques, such as non-equilibrium bosonization\cite{Levkivskyi16} for the anyonic statistics detection scenarios.\cite{rosenow16} Here, we present a fully analytical fourth-order perturbative calculation, 
leading to explicit formulas
for the current-current correlations at the outputs.


The structure of the paper is as follows. We start in section \ref{sec: system} with a description of the system and a short discussion of the quantum point contact geometries in FQHE.  In section \ref{sec: model}, we present our theoretical model of the quantum Hall edges and the two QPCs. In section \ref{sec: cross-corr}, \ref{sec: auto-corr} and \ref{sec: tunnel_current} we calculate the cross-correlations, the auto-correlations and the tunneling current of the setup, respectively. We generalize our results to finite temperature in section \ref{sec: finite_temperature}. We discuss the results of our theory in light of the recent experiment in section \ref{sec: discussion} and conclude in section \ref{sec: conclusion}.

\section{System description}
\label{sec: system}

\begin{figure}
    \includegraphics[scale=0.4]{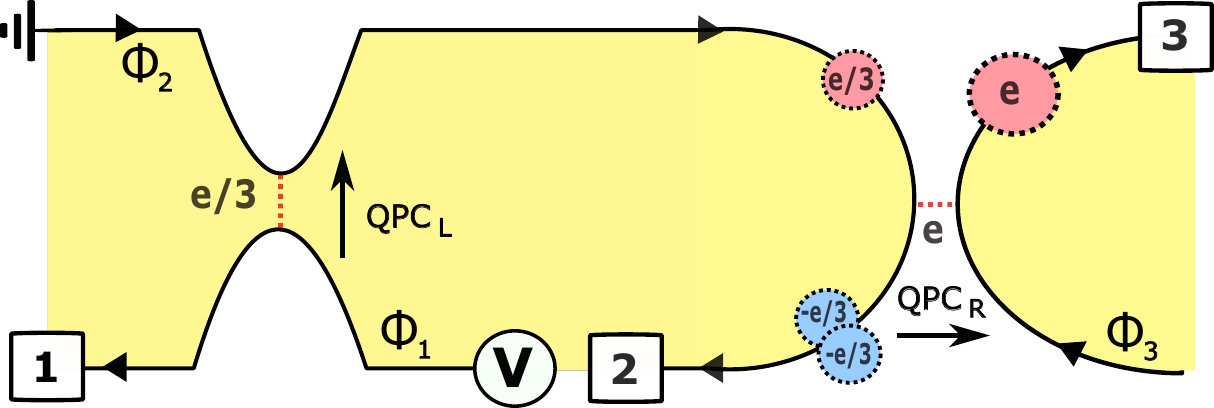}
   \caption{$\nu = 1/3$ fractional quantum Hall bar endowed with two QPCs. $QPC_L$ is placed in the weak-backscattering regime, emitting a dilute beam of $e/3$ quasiparticles on the edge $2$. These impinge on $QPC_R$ which is placed in the strong-backscattering regime, transmitting only electrons. The arrival of a $e/3$ quasiparticle on $QPC_R$ triggers the emission of an electron on edge $3$, and charge conservation implies that there must be a reflection of two $-e/3$ quasiholes on edge $2$. }
    \label{fig:setup}
\end{figure}

Our setup is shown schematically in Fig.~\ref{fig:setup}. A Hall bar is placed in the FQHE with a filling factor in the Laughlin series $\nu=1/m$ (the most practical case being $\nu=1/3$).
 The device is endowed with two distinct QPCs which can be individually controlled by voltages applied on corresponding split gates.

Let us first recall the two different regimes of a QPC, which are both used in this Andreev setup.
In the weak backscattering regime, the QPC brings the two edge states closer,
allowing for the tunneling of fractional quasiparticles of charge $e^*=e/m$, across the Hall
fluid, from one edge to the other. This is the configuration of $QPC_L$ in Fig.~\ref{fig:setup}.
 In this regime,  the voltage dependence of the tunneling current and the tunneling noise (i.e. zero-frequency fluctuations of the tunneling current) can be calculated theoretically using the perturbation theory of chiral Luttinger liquids. In the zero-temperature limit these are given by \cite{martin05}
    \begin{align}
        \left\langle I_T \right\rangle &= \frac{e^*~\Gamma_L^2}{v_F^{2\nu} \Gamma(2\nu)}\text{sgn}(V)\left( \frac{e^*V}{2\pi}\right)^{2\nu -1} , \\
        S_T  & = \frac{2e^{*2}~\Gamma_L^2}{v_F^{2\nu}  \Gamma(2\nu)}\left( \frac{e^*V}{2\pi}\right)^{2\nu -1} ,
    \label{eq:ITST}
  \end{align}
where $\Gamma_L$ is the (small) tunneling amplitude at $QPC_L$ and $v_F$ is the Fermi velocity along the edge.
The current and noise yield the relation
\begin{equation}
    S_T = 2e^*\left|\left\langle I_T \right\rangle \right| ,
    \label{fano_factor}
\end{equation}
which allows a diagnosis of the tunneling charge $e^*$. 

The opposite regime of a QPC is the strong coupling regime, where the effect of the QPC is strong enough to break the Hall fluid in two parts. It can be described perturbatively by
the tunneling of electrons through the vacuum between the two distinct parts. It is
the configuration of the right QPC on Fig.~\ref{fig:setup}, giving tunneling current and noise similar to
Eq.~\eqref{eq:ITST}, with $2 \nu$ replaced by $2/\nu$, and $e^*$ by $e$, yielding a relation
\begin{equation}
    S_T = 2e\left|\left\langle I_T \right\rangle \right| .
    \label{fano_factor_e}
\end{equation}

In the setup of Fig.~\ref{fig:setup}, the application of a positive voltage $V$ on the lower arm of $QPC_L$ leads to the emission of a dilute stream of $e^*$ quasiparticles on the edge $2$. It hence functions as a source of $e^*$ quasiparticles which impinge on the $QPC_R$, which is tuned to be in the strong-backscattering regime and hence transmits only electrons. It is the mismatch between the charge $e^*$ of the incoming excitations, and
the charge $e$ which can tunnel at $QPC_R$ which leads to quasiparticle Andreev reflection. When
the arrival of the $e^*$ quasiparticle triggers an electron tunnelling, 
charge conservation implies that two quasiholes (of charge $-e^*$ each) 
are reflected at $QPC_R$ (see Fig.~\ref{fig:setup}).
 Therefore, the transmitted current $I_3$ measured on contact $3$, and the reflected current $I_2$, measured on contact $2$, contain precious information about the quasiparticle Andreev reflection processes. 
 Of primary interest are the cross-correlations between currents on contacts 2 and 3, $\left<I_2 I_3 \right>$, the auto-correlations of the current at contact 3, $\left<I_3 I_3\right>$, and the average output current on contact 3, $\left<I_3 \right>$. In the Poissonian limit, these are expected to bear the following relations at zero temperature
\begin{align}
     \left<I_3 I_3 \right> &= 2e \left| \left\langle I_3 \right\rangle \right| , \\
     \left<I_2 I_3 \right> &= -\frac{4}{3}e \left|\left\langle I_3 \right\rangle\right| .
\end{align}
The first of these relations simply expresses that only electron tunneling contributes to 
$\langle I_3 \rangle$. The second relation expresses that for each electron tunneling contributing
to $\langle I_3 \rangle$, there is necessarily a charge $-2e/3$ which is reflected and which contributes to $\langle I_2 \rangle$, leading to a Fano factor of $(-2/3) \times 1$
for the cross-correlation.

It is important to stress the difference between this setup, and a similar setup where
both QPCs are placed in the weak backscattering regime, for example in the 
\emph{anyon collider}\cite{Ruelle2022}. When  $QPC_R$ is also in the weak coupling regime, 
and thus transmits $e^*$ quasiparticles,
it was shown that anyonic statistics plays an essential role in the tunneling current. The dominant contribution comes from interference between the process where the incoming $e^*$  reaches
the QPC before a thermal quasiparticle/quasihole pair is excited, and the
process where the order of these events is reversed. Because of the non-trivial anyonic
statistics between two $e^*$ quasiparticles, these interference effects have a finite contribution, which depends non-perturbatively
on the statistical properties of the incoming stream of $e^*$ quasiparticles \cite{Han2016,Morel22,Lee19}.

In the setup we are considering, since it is electrons that tunnel at $QPC_R$, the
exchange phase between the incoming $e^*$ quasiparticle and an electron (or a hole) is trivially
a multiple of $\pi$, and there is no contribution related to anyonic statistics.
The physics of Andreev reflection is thus fundamentally different from that of anyonic braiding.

\section{Theoretical model}
\label{sec: model}

We consider a fractional quantum Hall (FQH) bar in the Laughlin filling fraction. The quantum Hall bar is equipped with two quantum point contacts (QPC), denoted $QPC_L$ and $QPC_R$, as shown in Fig.~\ref{fig:setup}. 
The quantum Hall edges are modeled by chiral Luttinger liquids $\phi_j(x), j = 1, 2, 3$. The free Hamiltonian of this system is given by\cite{martin05, wen95}
\begin{equation}
    H_0 = \frac{v_F}{4\pi} \sum_{j=1}^{3} \int dx \left[\partial_{x}\phi_j(x) \right]^2 ,
\end{equation}
where $v_F$ is the Fermi velocity along the edges. Each edge carries a coordinate system $x$. $QPC_L$ operates in the weak-backscattering regime, allowing $e/m$ quasiparticles to tunnel across it. We apply a constant voltage $V$ on the edge $1$, causing $e/m$ quasiparticles to tunnel across $QPC_L$, from edge $1$ to edge $2$. This process is modeled by the tunneling Hamiltonian
\begin{equation}
    H_L(t) = \frac{\Gamma_L}{(2\pi a)^{\nu}} \sum_{\epsilon = \pm}O_L^\epsilon(t)e^{-{i \epsilon e^* V t}} ,
\end{equation}
where 
\begin{equation}
O_L^\pm(t) = F_L e^{\pm i \sqrt{\nu}\left[\phi_1(0,t) - \phi_2(0,t)\right]} ,
\label{eq:OL}
\end{equation}
$a$ is a short cutoff parameter, $e^* = e/m$ and $F_L$ is a Klein factor. The $e/m$ quasiparticles then travel along edge $2$ and are incident on $QPC_R$. We assume $e^*V \gg k_B T$,
 allowing us to work in the zero-temperature limit
(the finite temperature generalization is given in
Sec.~\ref{sec: finite_temperature}).
$QPC_R$ operates in the strong backscattering regime, allowing only electrons to tunnel between the two FQHE fluids. The tunneling of electrons across $QPC_R$ is given by
\begin{equation}
    H_R(t) = \frac{\Gamma_R}{(2\pi a)^{1/\nu}} \sum_{\epsilon = \pm}O_R^\epsilon(t) ,
\end{equation}
where 
\begin{equation}
O_R^\pm(t) = F_R \,  e^{\pm \frac{i} {\sqrt{\nu}}\left[\phi_2(L,t) - \phi_3(L,t)\right]} .
\label{eq:OR}
\end{equation}
The tunneling current operator across $QPC_R$ is readily obtained from the tunneling Hamiltonian and reads
\begin{equation}
        I_R(t) = i e \frac{\Gamma_R}{(2\pi a)^{1/\nu}} \sum_{\epsilon = \pm}\epsilon ~O_R^\epsilon(t) .
        \label{eq:IR}
\end{equation}
The Klein factors $F_L$, $F_R$ ensure that the tunneling operators
$H_L$ and $H_R$ commute, with $F_i F^{\dagger}_i= F^{\dagger}_i F_i =1$ ($i=L,R$) and $F_L F^{\dagger}_R = - F^{\dagger}_R F_L$.

The full Hamiltonian of the system is then given by
\begin{equation}
        H(t) = H_0 + H_T(t) ,
\end{equation}
where $H_T(t) = H_L(t) + H_R(t) $. Finally, the current operators on the edges are given by
\begin{equation}
    I_j(x,t) = \frac{\sqrt{\nu}e v_F}{2\pi}\partial_x\phi_j(x,t) \equiv \frac{e v_F}{2\pi i}\partial_x\partial_{\gamma}e^{i\gamma\sqrt{\nu}\phi_j(x,t)} \Big|_{\gamma = 0} ,
\end{equation}
where we introduced a dummy variable $\gamma$ , allowing us to express the current as
exponentials of the bosonic fields.

As explained in the previous section, there is a fundamental difference between the system that we are considering here, and a similar system where
the second QPC is also in the weak backscattering regime, used to
probe the anyonic statistics of the quasiparticles.\cite{rosenow16,bartolomei20,Morel22} In the latter
case, the two tunneling operators at the two QPCs have the same coefficient
$\sqrt{\nu}$ in the exponent [see Eq.~\eqref{eq:OL}],
which implies that the exchange of two quasiparticles
emitted at the two QPCs is accompanied by an anyonic phase $\pi \nu$. 
In the system that we consider here, the coefficient in the exponent of the left and right QPC are respectively $\sqrt{\nu}$ 
and $1/\sqrt{\nu}$ [as can be seen from Eqs.~\eqref{eq:OL} and \eqref{eq:OR}], which implies that the exchange of two quasiparticles emitted at the two QPC is accompanied by a trivial phase $\pi$.  For the case of the non-trivial anyonic exchange
phase, the perturbative calculation breaks down, and
other techniques, like non-equilibrium bosonization or resummation have
to be used.\cite{rosenow16, Morel22} Here, thanks to the trivial
exchange phase, the perturbative approach that we are using is valid,
as the higher orders terms in $\Gamma_L$, $\Gamma_R$ 
can be neglected in the limit of small $\Gamma_L$, $\Gamma_R$.

\section{Cross-correlations}
\label{sec: cross-corr}

\noindent
To capture the physics of quasiparticle Andreev reflection, we start by computing the cross-correlation noise of edges 2 and 3. This can be expressed using the Keldysh formalism as
\begin{equation}
    S_{23} = 2\int dt \left< T_K \left\{ \delta I_2(x,t^-) \delta I_3(y,0^+) e^{-i\int_K dt H_T(t)}\right\} \right>_C ,
\end{equation}
where $\delta I_j = I_j - \langle I_j \rangle$ and we only keep the "connected" contributions (denoted by the subscript $C$). Here $T_K$ denotes Keldysh time-ordering. We perform an expansion in the tunneling Hamiltonian up to the lowest non-trivial order $\Gamma_L^2\Gamma_R^2$ yielding
\begin{widetext}
\begin{align}
        S_{23} &= \frac{2\Gamma_L^2\Gamma_R^2}{(2\pi a)^{2\nu + \frac{2}{
    \nu}}} \int dt dt_1 \dots dt_4 \sum_{\{ \eta_k \}} \eta_1 \dots \eta_4 ~ e^{ie^*V (t_3 - t_4)} 
    \Big< T_K \Big\{\frac{e^2 v_F^2}{(2\pi i)^2}\partial_x\partial_{\gamma_1} e^{i\gamma_1\sqrt{\nu}\phi_2(x,t^-)} \partial_y\partial_{\gamma_2} e^{i\gamma_2\sqrt{\nu}\phi_3(y,0^+)} \nonumber \\ 
& \times e^{\frac{i}{\sqrt{\nu}}\left[\phi_2(L, t_1^{\eta_1}) - \phi_3(L, t_1^{\eta_1}) \right]}e^{-\frac{i}{\sqrt{\nu}}\left[\phi_2(L, t_2^{\eta_2}) - \phi_3(L, t_2^{\eta_2}) \right]} 
 e^{-i\sqrt{\nu}\left[\phi_1(0, t_3^{\eta_3}) - \phi_2(0, t_3^{\eta_3}) \right]} e^{i\sqrt{\nu}\left[\phi_1(0, t_4^{\eta_4}) - \phi_2(0, t_4^{\eta_4}) \right]}\Big\} \Big>_C \Big|_{\gamma_1,\gamma_2 = 0} ,
\end{align}
where $\eta_k = \pm $ indicates the Keldysh contour label, for the $k$-th time variable. Here, we do not write explicitly the Klein factors in the expression for the noise, as we consider a regime where they play no role in the end, see the discussion below Eq.~\eqref{eq:Kzerotemp} for a proper justification.

Correlation functions of multiple tunneling operators can be evaluated using the following identity
\begin{equation}
    \left<T_K\left\{ \prod_{j=1}^{N}e^{i\alpha_j\phi_\mu \left( x_j,t_j^{\eta_j}\right)} \right\} \right> = \text{exp}\left[ \sum_{j=1}^{N}\sum_{k=j+1}^{N}\alpha_j\alpha_k~\mathcal{G}^{\eta_{j}\eta_{k}}(x_j-x_k, t_j-t_k)\right] ,
    \label{wick_theorem}
\end{equation}  
which is non-zero only when $\sum_{j}\alpha_j = 0$, allowing us to express $S_{23}$ as
\begin{align}
S_{23} &= \frac{2e^2 v_F^2}{(2\pi i)^2}\frac{\Gamma_L^2\Gamma_R^2}{(2\pi a)^{2\nu + 2/
    \nu}} \int dt dt_1 \dots dt_4 \sum_{\{ \eta_k \}} \eta_1 \dots \eta_4 ~ e^{ie^* V t_{34}} e^{2\nu\mathcal{G}^{\eta_{3}\eta_{4}}(0, t_{34})}  e^{\frac{2}{\nu}\mathcal{G}^{\eta_{1}\eta_{2}}(0, t_{12})} \nonumber  \\  
& \times \partial_x\partial_{\gamma_1} \left[ e^{-\gamma_1\mathcal{G}^{-\eta_1}(x-L,t-t_1)}e^{\gamma_1\mathcal{G}^{-\eta_2}(x-L,t-t_2)} e^{-\gamma_1\nu\mathcal{G}^{-\eta_3}(x,t-t_3)}e^{\gamma_1\nu\mathcal{G}^{-\eta_4}(x,t-t_4)} \right]_{\gamma_1=0} \nonumber \\
& \times \partial_y\partial_{\gamma_2}\left[e^{\gamma_2\mathcal{G}^{+\eta_1}(y-L,-t_1)}e^{-\gamma_2\mathcal{G}^{+\eta_2}(y-L,-t_2)} \right]_{\gamma_2 = 0}
\left[ e^{-\mathcal{G}^{\eta_{1}\eta_{3}}(L,t_{13})}e^{\mathcal{G}^{\eta_{1}\eta_{4}}(L,t_{14})}e^{\mathcal{G}^{\eta_{2}\eta_{3}}(L,t_{23})}e^{-\mathcal{G}^{\eta_{2}\eta_{4}}(L,t_{24})}  - 1\right] .
\label{cross_corr_intermediate}
\end{align}
\end{widetext}
The final term in the last line of Eq.~\eqref{cross_corr_intermediate} is conveniently rewritten as $K_{1234} -1$. The $-1$ contribution removes all the disconnected diagrams up to order $\Gamma_L^2\Gamma_R^2$, while the term $K_{1234}$ involves two-point correlators of bosonic fields, with each of the two bosonic fields coming from the tunneling operators at different QPCs. This term is given by:
\begin{equation}
    K_{1234} = \frac{\exp \left[ \mathcal{G}^{\eta_1\eta_4}(L, t_{14})+\mathcal{G}^{\eta_2\eta_3}(L, t_{23}) \right]}{\exp \left[ \mathcal{G}^{\eta_1\eta_3}(L, t_{13})+\mathcal{G}^{\eta_2\eta_4}(L, t_{24}) \right]} .
\end{equation}
The fact that here, the prefactor of each Green's function in the exponential is equal to 1 is because of the coefficients $\sqrt{\nu}$ and $1/\sqrt{\nu}$ of the tunneling operators of the left and right QPC respectively,
whose product is 1.  It is thus a consequence of the  trivial exchange phase between the excitations of the two QPC, which allows a simpler analytic treatment.

Here $\mathcal{G}^{\eta_i \eta_j}(x, \tau)$ is the chiral Luttinger liquid Green's function with the superscripts corresponding to the Keldysh time ordering, and we used the shorthand notation $t_{ij} = t_i - t_j$. At zero temperature, the Green´s function is given by
\begin{equation}
    \mathcal{G}^{\eta_{1}\eta_{2}}(x,\tau) = \log \left(\frac{\tau_0}{\tau_0 + i (\tau - x/v_F) \chi_{12}(\tau)}\right)  ,
\label{zero_temp_green_fn}
\end{equation}
where $\tau_0 = a/v_F$ and 
\begin{equation}
    \chi_{12}(\tau) = \frac{1}{2}\left[ (\eta_2 - \eta_1) + \text{sgn}(\tau)(\eta_1 + \eta_2)\right] .
\label{keldysh_index} 
\end{equation}

In the following, we consider the limit of a large separation
$L$ between the two QPCs, effectively taking $L \to \infty$.
Hence, the tunneling times $t_{1/2}$ at the right QPC are
always much greater than the tunneling times $t_{3/4}$ at the left QPC. This allows us to simplify  Eq.~\eqref{keldysh_index} and directly write $\chi_{13}(t_{13}) = \chi_{23}(t_{23}) = \eta_3$ and $\chi_{14}(t_{14}) = \chi_{24}(t_{24}) = \eta_4$. This in turn enables to greatly simplify the term $K_{1234}$ which becomes, at zero temperature:
\begin{equation}
K_{1234} = \frac{\left[t_1 - t_3 -\frac{L}{v_F} - i\eta_3 \tau_0 \right]\left[ t_2 - t_4 -\frac{L}{v_F} - i\eta_4 \tau_0 \right]}{\left[t_1 - t_4 -\frac{L}{v_F}- i\eta_4 \tau_0 \right] \left[t_2 - t_3 -\frac{L}{v_F}-i\eta_3 \tau_0 \right]}   .
\label{eq:Kzerotemp}
\end{equation}
Note that it is this same assumption $L \to \infty$ which
allows us to ignore the Klein factors in the calculation.
Indeed $t_{1,2} \gg t_{3,4}$ 
implies that the two tunneling operators at the right QPC (with times 
$t_1$ and $t_2$), are always well separated from those at the left QPC (with times $t_3$ and $t_4$)
on the Keldysh contour, so that the contribution from the Klein
factors always reduces to $F_L F^{\dagger}_L \times 
F_R F^{\dagger}_R =1 $.

From Eq.~\eqref{zero_temp_green_fn}, one readily sees that the derivative of the Green's function at zero temperature can be expressed as a delta function in the limit of vanishingly small cutoff
\begin{align}
    &\partial_x\mathcal{G}^{\eta_1\eta_2}(x-y,t_1-t_2) \nonumber \\
    &= \begin{cases}
                        +\frac{i \pi}{v_F}\eta_2 \delta \left(t_1-t_2 - \frac{x-y}{v_F} \right), ~ x-y > 0 \\
                        -\frac{i \pi}{v_F}\eta_1 \delta \left(t_1-t_2 - \frac{x-y}{v_F} \right), ~ x-y < 0
                    \end{cases}    
\label{green_fn_derivative}                    
\end{align}
After making the transformation $t_{1/2} \to t_{1/2} + L$, and taking the limit $L\to\infty$, one can further simplify the expression for the cross-correlations using the above relation, Eq.~\eqref{green_fn_derivative}. This enables us to evaluate the integrals over $t$ and $t_1$, and to get rid of the delta functions to obtain:
\begin{align}
    S_{23} &= \frac{e^2\Gamma_L^2\Gamma_R^2}{2 (2\pi a)^{2\nu + \frac{2}{\nu}}} \int dt_2  dt_3 dt_4 \sum_{\{\eta_k\}}\eta_1\dots\eta_4 e^{ie^*Vt_{34}} \nonumber \\
    & \times \left[(\eta_2 - \eta_1) + \nu(\eta_4 - \eta_3) \right]  e^{2\nu\mathcal{G}^{\eta_3\eta_4}(0,t_{34})}  e^{\frac{2}{\nu}\mathcal{G}^{\eta_1\eta_2}(0,-t_{2})} \nonumber \\ 
    & \times \left(\eta_1  - \eta_2 \right) \left\{ \frac{\left[ - t_3  - i\eta_3 \tau_0 \right]\left[ t_{2} - t_4  - i\eta_4 \tau_0 \right]}{\left[ - t_4 - i\eta_4 \tau_0 \right] \left[t_{2} - t_3 -i\eta_3 \tau_0 \right]} - 1 \right\} .
    \label{cross_corr_three_integrals}
\end{align}

Further manipulations and the coordinate transformation $t_3 - t_4 = \tau$, $(t_3 + t_4)/2 = T$ allow us to express the cross-correlation as
\begin{align}
S_{23} &= \frac{e^2\Gamma_L^2\Gamma_R^2}{ (2\pi a)^{2\nu + \frac{2}{\nu}}}\int dt_2 d\tau e^{ie^*V \tau} \sum_{\{\eta_k\}}  e^{\frac{2}{\nu}\mathcal{G}^{\eta_1\eta_2}(0,-t_{2})}   \nonumber \\
& \times \left[\eta_3\eta_4(1- \eta_1 \eta_2) + \nu \frac{\eta_1 - \eta_2}{2}(\eta_4 - \eta_3) \right] e^{2\nu\mathcal{G}^{\eta_3\eta_4}(0,\tau)} \nonumber \\ 
&\times \int dT \frac{-t_2 \left[\tau + i (\eta_3 - \eta_4) \tau_0 \right]}{\left[ - T + \frac{\tau}{2} - i\eta_4 \tau_0 \right] \left[t_{2} - T - \frac{\tau}{2} -i \eta_3 \tau_0 \right]} .
\label{cross_corr_first_piece1}
\end{align}

The $T$-integral has been evaluated in Appendix \ref{app:contour} using Cauchy's residue theorem. Using this result and subsequently summing over the Keldysh indices $\eta_2$ and  $\eta_4$, this leads to
\begin{align}
S_{23} &= \frac{4 i \pi e^2\Gamma_L^2\Gamma_R^2}{ (2\pi a)^{2\nu + \frac{2}{\nu}}}\int dt_2 d\tau e^{ie^*V \tau} \sum_{\eta_3} e^{2\nu\mathcal{G}^{\eta_3,-\eta_3}(0,\tau)}    \nonumber \\
& \times  \sum_{\eta_1} (\eta_3 + \nu \eta_1) \left( \frac{\tau_0}{\tau_0 + i \eta_1 t_2}
\right)^{2/\nu}
 \frac{t_2 \left(\tau + 2 i \eta_3 \tau_0 \right)}{ - t_{2} + \tau + 2 i \eta_3 \tau_0} .
\label{cross_corr_first_piece2}
\end{align}

The $t_2-$integrals have been evaluated in Appendix \ref{app:contour}. Subsequent trivial manipulations give us, to leading order in the cutoff,
\begin{align}
S_{23} &= - \frac{4  e^2\Gamma_L^2\Gamma_R^2}{v_F^2 (2\pi a)^{2\nu + \frac{2}{\nu}-2}}   (1-\nu)       \nonumber \\
& \qquad  \times \int  d\tau \cos \left( e^*V \tau \right)    \left( \frac{\tau_0}{\tau_0 - i \tau} \right)^{2 \nu + 2/\nu - 2} .
\label{cross_corr_tau_integral}
\end{align}
Finally, the $\tau-$integral can be evaluated using known results \cite{gradshteyn14} to give
\begin{align}
S_{23} &= - (1-\nu)  \frac{ 2 e^2\Gamma_L^2\Gamma_R^2}{v_F^3 \Gamma \left( 2 \nu + \frac{2}{\nu} -2 \right)}          \left| \frac{e^* V}{2 \pi v_F} \right|^{2 \nu + \frac{2}{\nu} -3} ,
\label{eq:final_crosscorr}
\end{align}
where the function $\Gamma(x)$ in the denominator denotes the Euler-Gamma function. 

\section{Auto-correlations}
\label{sec: auto-corr}

Now we go on to compute the auto-correlation noise on edge 3, which is expressed using the Keldysh formalism as
\begin{equation}
    S_{33} = 2\int dt \left< T_K \left\{ \delta I_3(x,t^-) \delta I_3(y,0^+)e^{-i\int_K dt H_T(t)}\right\} \right>_C .
\end{equation}

Performing again an expansion in the tunneling Hamiltonian up to order $\Gamma_L^2\Gamma_R^2$ yields
\begin{align}
&S_{33} = \frac{2e^2 v_F^2\Gamma_L^2\Gamma_R^2}{(2\pi a)^{2\nu + \frac{2}{
    \nu}}} \int dt dt_1 \dots dt_4 \sum_{\{ \eta_k \}} \eta_1 \dots \eta_4 ~ e^{ie^*Vt_{34}} \nonumber \\
&\Big< T_K \Big\{\frac{1}{(2\pi i)^2}\partial_x\partial_{\gamma_1} e^{i\gamma_1\sqrt{\nu}\phi_3(x,t^-)} \partial_y\partial_{\gamma_2} e^{i\gamma_2\sqrt{\nu}\phi_3(y,0^+)} \nonumber \\ 
& 
e^{-i\sqrt{\nu}\left[\phi_1(0, t_3^{\eta_3}) - \phi_2(0, t_3^{\eta_3}) \right]} 
e^{-\frac{i}{\sqrt{\nu}}\left[\phi_2(L,t_2^{\eta_2 }) - \phi_3(L, t_2^{\eta_2}) \right]} \nonumber\\
& 
e^{\frac{i}{\sqrt{\nu}}\left[\phi_2(L, t_1^{\eta_1}) - \phi_3(L, t_1^{\eta_1}) \right]}
e^{i\sqrt{\nu}\left[\phi_1(0, t_4^{\eta_4}) - \phi_2(0, t_4^{\eta_4}) \right]}\Big\} \Big>_C \Big|_{\gamma_1,\gamma_2 = 0} ,
\label{eq:auto_corr_intermediate}
\end{align}
where we are only interested in the zero-temperature regime at this stage, thus allowing us to discard lower order terms corresponding to purely thermal contributions to the noise.

Proceeding along the same lines as the derivation of the cross-correlations, and
employing then Eqs.~\eqref{wick_theorem} and \eqref{green_fn_derivative} one obtains:
\begin{align}
S_{33} &= \frac{e^2\Gamma_L^2\Gamma_R^2}{2(2\pi a)^{2\nu + \frac{2}{
    \nu}}} \int dt_2 dt_3 dt_4 \sum_{\{ \eta_k \}} \eta_1 \dots \eta_4  \left(\eta_1 -\eta_2 \right)^2 \nonumber \\ 
& \times e^{ie^*Vt_{34}} e^{2\nu\mathcal{G}^{\eta_3\eta_4}(0,t_{34})} e^{\frac{2}{\nu}\mathcal{G}^{\eta_1\eta_2}(0,-t_{2})} \nonumber \\
& \times \left\{ \frac{\left[ - t_3 - i\eta_3 \tau_0 \right]\left[ t_2 - t_4  - i\eta_4 \tau_0 \right]}{\left[ - t_4 - i\eta_4 \tau_0 \right] \left[t_2 - t_3 -i\eta_3 \tau_0 \right]} 
 - 1\right\} .
 \label{auto_corr_three_integrals}
\end{align}
Comparing Eq.~\eqref{auto_corr_three_integrals} with Eq.~\eqref{cross_corr_three_integrals}, one finds that the expression for the auto-correlations is the same as the first piece of the cross-correlations, up to a minus sign. Hence, the calculations of the cross-correlations follow through for auto-correlations as well, giving us finally
\begin{align}
S_{33} &= \frac{ 2 e^2\Gamma_L^2\Gamma_R^2}{v_F^3 \Gamma \left( 2 \nu + \frac{2}{\nu} -2 \right)}          \left| \frac{e^* V}{2 \pi v_F} \right|^{2 \nu + \frac{2}{\nu} -3} .
\label{eq:final_autocorr}
\end{align}

\section{Tunneling current}
\label{sec: tunnel_current}

Now we calculate the average tunneling current across $QPC_R$ which is expressed in the Keldysh formalism as
\begin{equation}
    \left< I_3 (x, t)\right> =  \left< T_K \left\{ \frac{1}{2}\sum_{\eta_1}I_R(t_1^{\eta_1})~e^{-i\int_K dt H_T(t)}\right\} \right> ,
\end{equation}
where $t_1 = t - \frac{x-L}{v_F}$ and the tunneling current operator $I_R$ was defined in Eq.~\eqref{eq:IR}. 
Expanding in the tunneling Hamiltonian up to third order and retaining only the relevant, non-zero, order $\Gamma_L^2\Gamma_R^2$ terms, one is left with
\begin{align}
         \left< I_3 (x,t) \right> &=-\frac{e\Gamma_L^2\Gamma_R^2}{4 (2\pi a)^{2\nu + \frac{2}{\nu}-2}}\int dt_2dt_3dt_4 \nonumber \\
         & \times \sum_{\{\eta_k,\epsilon_k\}}\epsilon_1\eta_2\eta_3\eta_4 e^{- i e^* V (\epsilon_3 t_3 + \epsilon_4 t_4)} \nonumber \\
         & \times \left< T_K \left\{ O_R^{\epsilon_1}(t_1^{\eta_1}) O_R^{\epsilon_2}(t_2^{\eta_2}) O_L^{\epsilon_3}(t_3^{\eta_3}) O_L^{\epsilon_4}(t_4^{\eta_4})   \right\} \right>_C .
         \label{eq:current_intermediate}
\end{align}
The $\epsilon_k$ indices are then summed over, and the resulting Keldysh correlators are subsequently evaluated using Wicks theorem [see eq.~\eqref{wick_theorem}] giving us
\begin{align}
        \left< I_3\right> &= i \frac{e \Gamma_R^2\Gamma_L^2}{(2\pi a)^{2\nu + 2/\nu}}\sum_{\{\eta_k \}} \eta_2\eta_3\eta_4 
        \int dt_2 d\tau \sin (e^*V\tau) \nonumber \\
        & \times e^{2\nu\mathcal{G}^{\eta_3\eta_4}(\tau)}  e^{\frac{2}{\nu}\mathcal{G}^{\eta_1\eta_2}(-t_{2})} \nonumber \\
        & \times \int dT \frac{-t_2 \left[\tau + i (\eta_3 - \eta_4) \tau_0 \right]}{\left[ - T + \frac{\tau}{2} - i\eta_4 \tau_0 \right] \left[t_{2} - T - \frac{\tau}{2} -i \eta_3 \tau_0 \right]} .
\end{align}
Noting that a proper rescaling of the integrated variables $t_2 \to t_2 + t_1$ and $T \to T + t_1$ allows us to get rid of the external variable $t_1$, we argue that the tunneling current is constant and does not depend on $x,t$.

Performing the summation over Keldysh indices $\eta_1$ and $\eta_2$, the integrals over $T$ and $t_2$ are then carried out similarly to the previous sections, relying on the results of Appendix \ref{app:contour} yielding
%
%
\begin{align}
        \left< I_3\right> &= - 2i (2 \pi)^2 \frac{e \Gamma_R^2\Gamma_L^2}{(2\pi a)^{2\nu + 2/\nu}} \tau_0^2
        \nonumber \\
        & \qquad \times \int  d\tau \sin (e^*V\tau)  \left( \frac{\tau_0}{\tau_0 - i \tau}\right)^{2\nu + 2/\nu -2} . 
\end{align}
Performing the final $\tau-$integral, one is left with
\begin{align}
        \left< I_3\right> &=   \frac{e \Gamma_R^2\Gamma_L^2}{v_F^3 \Gamma \left( 2\nu + \frac{2}{\nu} -2 \right)}  
         \left| \frac{e^* V}{2 \pi v_F} \right|^{2\nu + 2/\nu -3}  \text{Sgn}(V) .
\label{eq:finalcurrent}         
\end{align}

\section{Finite temperature}
\label{sec: finite_temperature}

In this section, we generalize the results presented in the last three sections to finite temperatures. We start by noting that the expression of cross-correlations in Eq.~\eqref{cross_corr_intermediate}, along with the one for the tunneling current in Eq.~\eqref{eq:current_intermediate}, both apply to a general temperature, provided that one uses the finite temperature Green's function. Similarly, the expression of Eq.~\eqref{eq:auto_corr_intermediate} for the auto-correlations is also valid at finite temperature but it now only describes the "excess" auto-correlations $S_{33}^\text{exc} = S_{33} (V) - S_{33} (V=0)$, as we are not interested in the purely thermal contribution.

Substituting then, instead of the zero-temperature Green's function, the finite-temperature one 
    \begin{equation}
          \mathcal{G}^{\eta_{1}\eta_{2}}(x,\tau) = \log \left(\frac{\sinh (i\pi \theta \tau_0)}{\sinh \left[ \pi \theta \left(i\tau_0 - \chi_{12}(\tau)(\tau-x/v_F)\right)\right]}\right) ,
\label{finite_temp_green_fn}
    \end{equation}
gives us the corresponding generalization to finite temperature $\theta$. It is important to stress out that, while these can be generalized, all the calculations presented in the appendices, and used in the previous sections, are valid only at zero temperature.

While the finite temperature extension does not present any formal difficulties, it is still a long and rather tedious calculation. For clarity sake, we choose to present only the final results here, the main steps of the derivation being presented in Appendix \ref{app:finiteT}. The tunneling current at finite temperature $\theta$ is given by
\begin{align}
   \left< I_3\right> = &   \frac{e \Gamma_R^2\Gamma_L^2}{\pi v_F^3 \Gamma \left( 2\nu + \frac{2}{\nu} -2 \right)}  
         \left( \frac{\theta}{v_F} \right)^{2\nu + 2/\nu -3} \sinh \left( \frac{e^* V}{2 \theta}\right) \nonumber \\
         & \times \left| \Gamma \left( \nu + \frac{1}{\nu} - 1 + i \frac{e^* V}{2 \pi \theta}\right) \right|^2  .  
\label{eq:currentfiniteT}
\end{align}

For $\nu = 1/3$, we obtain for the excess auto-correlations
\begin{widetext}
\begin{equation}
    S_{33}^\text{exc} = \frac{4 e^2\Gamma_L^2\Gamma_R^2}{\pi^2 v_F^3} \left( \frac{\theta}{v_F} \right)^{11/3}\sinh\left(\frac{eV}{6\theta}\right)\frac{\left| \Gamma\left(\frac{7}{3}+ i \frac{eV}{6\pi\theta} \right)\right|^2}{\Gamma\left(\frac{14}{3}\right)}\text{Im}\left[\psi\left(\frac{7}{3}+ i \frac{eV}{6\pi\theta} \right) \right] ,
 \label{eq:autocorrfiniteT}   
\end{equation}
while the cross-correlation can also be calculated similarly yielding
\begin{align}
        S_{23} = 
        - \frac{2}{3} \frac{4 e^2\Gamma_L^2\Gamma_R^2}{\pi^2 v_F^3} \left( \frac{\theta}{v_F} \right)^{11/3} &
         \left\{ \sinh\left(\frac{eV}{6\theta}\right) \frac{\left| \Gamma\left(\frac{7}{3}+ i \frac{eV}{6\pi\theta} \right)\right|^2}
        {\Gamma\left(\frac{14}{3}\right)} \text{Im}\left[\psi\left(\frac{7}{3} + i\frac{eV}{6\pi\theta}\right) \right]   \right. \nonumber \\
        & - 
        \left. \frac{1}{8} \frac{\left| \Gamma\left(\frac{1}{3}+ i \frac{eV}{6\pi\theta} \right)\right|^2}{\Gamma\left(\frac{5}{3}\right)} \frac{eV}{6\pi\theta}\sinh\left( \frac{eV}{6\theta}\right) \left[\frac{23}{120} + \frac{9}{160}\left( \frac{eV}{3\pi\theta}\right)^2  \right]\right\},
  \label{eq:crosscorrfiniteT}      
\end{align}
where $\psi$ is the digamma function. One can readily check that in the limit $\theta \longrightarrow 0$, these results reproduce the ones obtained in Eqs.~\eqref{eq:final_autocorr} and \eqref{eq:final_crosscorr}.
\end{widetext}

\section{Discussion}
\label{sec: discussion}

\begin{figure}
    \centering
\includegraphics[scale=0.55]{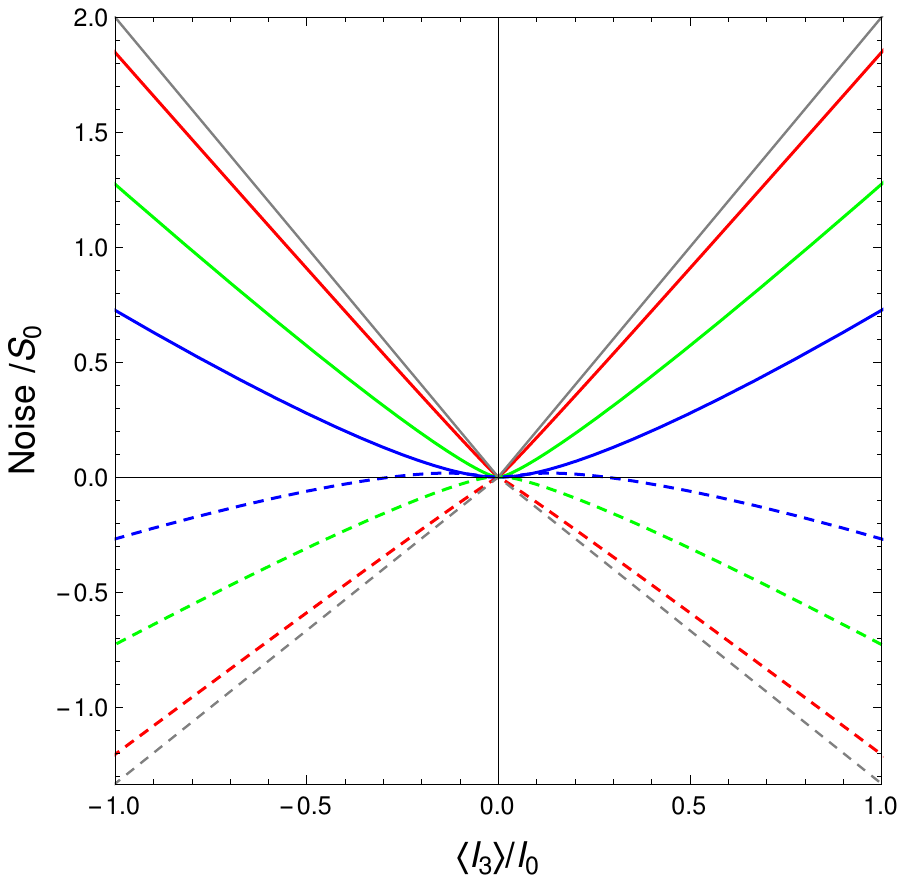}
    \caption{Auto-correlation noise $S_{33}^\text{exc}$ (full lines) and cross-correlation noise $S_{23}$ (dashed lines) plotted as a function of the tunneling current 
    $\langle I_3 \rangle$, for different values 
    of the temperature $k_B \theta = 0$ (gray), 0.1 (red), 0.5 (green), 1 (blue) (arbitrary energy units), with $eV$ varying from -32 to +32 in the same units. The zero-temperature case (gray lines) correspond to Eqs~(\ref{eq:final_crosscorr}), (\ref{eq:final_autocorr}) and (\ref{eq:finalcurrent}), while the finite temperature
    case correspond to Eqs.~(\ref{eq:currentfiniteT})-(\ref{eq:crosscorrfiniteT}).
    The current is normalized by $I_0 = \frac{e \Gamma_R^2\Gamma_L^2}{\pi v_F^3}$, 
    the noise by $S_0 = \frac{4 e^2 \Gamma_R^2\Gamma_L^2}{\pi^2 v_F^3}$.}
    \label{fig:noise-current-plot}
\end{figure}

Consolidating the calculations of auto-correlations, cross-correlations, and the tunneling current at zero temperature, we can see that they satisfy the following relations for $\nu = 1/3$
\begin{align}
\label{noise-current-reln1}
        S_{33} &= 2e\left|\left\langle I_3\right\rangle \right| , \\
        S_{23} &= -\frac{4}{3}e \left|\left\langle I_3\right\rangle\right| ,
        \label{noise-current-reln2}
\end{align}
which matches the experimental results of Ref.~\onlinecite{Glidic2023}. Eqs.~\eqref{noise-current-reln1} and \eqref{noise-current-reln2} have been plotted as the gray straight lines in Fig.~\ref{fig:noise-current-plot}.

The (-4/3) coefficient of the cross-correlation noise generalizes to $ -2 (m-1)/m$ for
an arbitrary Laughlin fraction $\nu=1/m$.
The ratio between the cross- and auto-correlations, equal to
$-2/3$ (or more generally $(m-1)/m$) is thus a direct manifestation of 
the Andreev reflection process, where the transmission of a charge
$e \times m/m$ (i.e. an electron) is always accompanied of the reflection of a negative charge 
$e \times (1-m)/m $ [i.e. $(m-1)$ holes of charge $-e/m$].

The analytical results at finite temperature allows
us to see how the current and noise are modified when the ratio $k_B \theta/ e V$ is non-zero.
This is first illustrated on Fig.~\ref{fig:noise-current-plot}, which in addition to the zero-temperature results shows the auto-correlations and cross-correlations noises as a function of the tunneling current $\langle I_3 \rangle$ for different
values of the temperature $\theta$. One can see that the finite temperature leads to a rounding
of the noise as function of the current and thus for a lower slope for the noise as function of the current until $e V \gg k_B \theta$ where the zero-temperature slope is recovered.

\begin{figure}[t]
    \centering
    \includegraphics[width=0.48\textwidth]{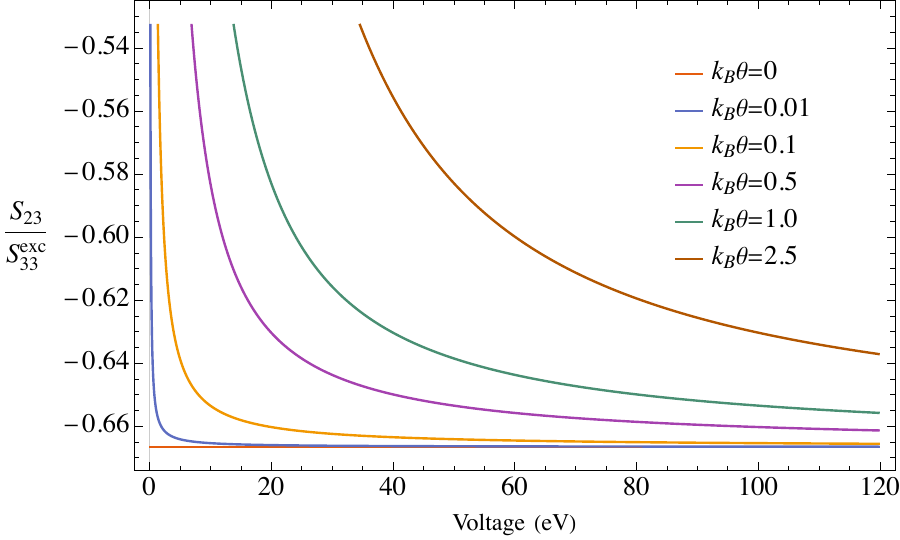}
    \includegraphics[width=0.48\textwidth]{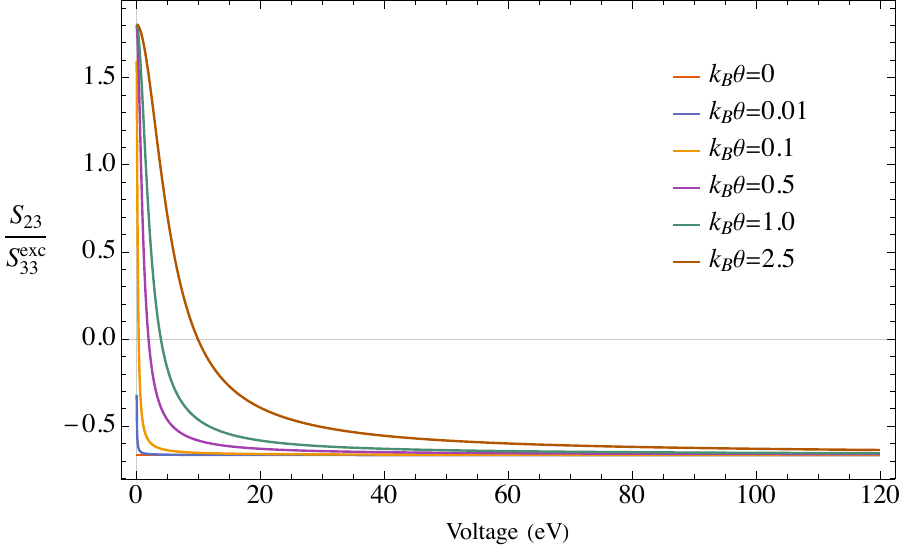}
   \caption{Ratio of cross-correlations to auto-correlations as a function of the voltage $e V$ (in arbitrary energy unit) for different values of $k_B \theta$ (in the same unit). The ratio asymptotically reaches down to $-2/3$ for $e V \gg k_B \theta$, in agreement with the zero temperature result. For smaller
   values of $e V/ k_B\theta$, the ratio increases, and becomes positive for small values
   of $e V/k_B \theta$.}
    \label{fig: plot}
\end{figure}

To examine in more details the impact of temperature on the ratio $S_{23}/S_{33}^\text{exc}$, Fig.~\ref{fig: plot} shows this ratio as a function of voltage at different temperatures. One can see that for large $e V \gg k_B \theta$, 
 the noise ratio always converge towards the
 zero-temperature value $S_{23}/S_{33}^\text{exc} = -2/3$. As $e V/k_B \theta$ decreases. the ratio 
 increases, and changes sign close to $e V/k_B \theta = 4$. It reaches
 a value $\sim1.8$ at $e V/k_B \theta=0$.
 
It is also possible to reverse the configuration of the QPCs such that $QPC_L$ emits electrons and $QPC_R$ transmit $e/3$ quasiparticles. In this case, we expect a similar process where a quasiparticle is transmitted across $QPC_R$ and two quasiparticles are Andreev reflected
(with here Andreev reflected charges of the same sign as the transmitted charge). This results in the relation $S_{23} = 2S_{33}$.

\section{Conclusions}
\label{sec: conclusion}

Motivated by recent experiments which demonstrate Andreev-like  reflection in the $\nu=1/3$ FQHE, we have presented an analytical calculation of the output tunneling current, cross-correlation noise, and the auto-correlation noise that follows exactly the measurement protocol of Ref.~\onlinecite{Glidic2023}. 
We have performed an exact perturbative calculation at order
four in the tunneling amplitudes of the two QPCs, 
beyond the usual second order calculations for the current and noise at a QPC. We have obtained
explicit analytical expressions for the current and noises, both
at zero temperature and at finite temperature.
Our results match that of the experiment, in
particular for the ratio of the crossed- and auto-correlation noises,
which is a direct manifestation of the Andreev reflection process.
Our results also show how finite temperature will modify the noise ratio,
which can even become positive when $V/\theta$ becomes small.


A natural extension of the present work would be to study Andreev reflection in non-Laughlin filling fractions of the FQHE. For a $\nu=2/5$ filling fraction for instance, a considerable challenge would be to include the fact that depending on the operating regimes, several types of co-propagating bosonic excitations can coexist at the location of the two QPCs\cite{shtanko14}, necessitating some choice for the proper theoretical model. Recent experimental results on anyon collisions in $\nu = 2/5$\cite{Ruelle2022, Glidic2022braiding, Lee2022}, allow us to exploit  experimental knowledge of the different operating regimes, and one should therefore be able to specify the types of quasiparticles which tunnel at the location of the two QPCs. Of interest is also $\nu=2/3$ which is known to involve two counter-propagating edge excitations, ultimately giving rise to neutral modes where energy -- rather than charge -- can propagate upstream from the quasiparticle charge current. 

\acknowledgments

We thank F. Pierre and A. Anthore for enlightning discussions on their experiment.
This work was carried out in the framework of the project
``ANY-HALL'' (Grant ANR No ANR-21-CE30-0064-03).
It received support from the French government under the France 2030 investment plan, as part of the Initiative d'Excellence d'Aix-Marseille Université - A*MIDEX.
We acknowledge support from the institutes IPhU (AMX-19-IET-008) and AMUtech (AMX-19-IET-01X).

\bibliography{citations}

\begin{thebibliography}{37}%
\makeatletter
\providecommand \@ifxundefined [1]{%
 \@ifx{#1\undefined}
}%
\providecommand \@ifnum [1]{%
 \ifnum #1\expandafter \@firstoftwo
 \else \expandafter \@secondoftwo
 \fi
}%
\providecommand \@ifx [1]{%
 \ifx #1\expandafter \@firstoftwo
 \else \expandafter \@secondoftwo
 \fi
}%
\providecommand \natexlab [1]{#1}%
\providecommand \enquote  [1]{``#1''}%
\providecommand \bibnamefont  [1]{#1}%
\providecommand \bibfnamefont [1]{#1}%
\providecommand \citenamefont [1]{#1}%
\providecommand \href@noop [0]{\@secondoftwo}%
\providecommand \href [0]{\begingroup \@sanitize@url \@href}%
\providecommand \@href[1]{\@@startlink{#1}\@@href}%
\providecommand \@@href[1]{\endgroup#1\@@endlink}%
\providecommand \@sanitize@url [0]{\catcode `\\12\catcode `\$12\catcode
  `\&12\catcode `\#12\catcode `\^12\catcode `\_12\catcode `\%12\relax}%
\providecommand \@@startlink[1]{}%
\providecommand \@@endlink[0]{}%
\providecommand \url  [0]{\begingroup\@sanitize@url \@url }%
\providecommand \@url [1]{\endgroup\@href {#1}{\urlprefix }}%
\providecommand \urlprefix  [0]{URL }%
\providecommand \Eprint [0]{\href }%
\providecommand \doibase [0]{http://dx.doi.org/}%
\providecommand \selectlanguage [0]{\@gobble}%
\providecommand \bibinfo  [0]{\@secondoftwo}%
\providecommand \bibfield  [0]{\@secondoftwo}%
\providecommand \translation [1]{[#1]}%
\providecommand \BibitemOpen [0]{}%
\providecommand \bibitemStop [0]{}%
\providecommand \bibitemNoStop [0]{.\EOS\space}%
\providecommand \EOS [0]{\spacefactor3000\relax}%
\providecommand \BibitemShut  [1]{\csname bibitem#1\endcsname}%
\let\auto@bib@innerbib\@empty
\bibitem [{\citenamefont {Tinkham}(1996)}]{tinkham96}%
  \BibitemOpen
  \bibfield  {author} {\bibinfo {author} {\bibfnamefont {M.}~\bibnamefont
  {Tinkham}},\ }\href {https://books.google.fr/books?id=XP\_uAAAAMAAJ} {\emph
  {\bibinfo {title} {Introduction to Superconductivity}}},\ International
  series in pure and applied physics\ (\bibinfo  {publisher} {McGraw Hill},\
  \bibinfo {year} {1996})\BibitemShut {NoStop}%
\bibitem [{\citenamefont {Cuevas}\ \emph {et~al.}(1996)\citenamefont {Cuevas},
  \citenamefont {Mart\'{\i}n-Rodero},\ and\ \citenamefont {Yeyati}}]{cuevas96}%
  \BibitemOpen
  \bibfield  {author} {\bibinfo {author} {\bibfnamefont {J.~C.}\ \bibnamefont
  {Cuevas}}, \bibinfo {author} {\bibfnamefont {A.}~\bibnamefont
  {Mart\'{\i}n-Rodero}}, \ and\ \bibinfo {author} {\bibfnamefont {A.~L.}\
  \bibnamefont {Yeyati}},\ }\href {\doibase 10.1103/PhysRevB.54.7366}
  {\bibfield  {journal} {\bibinfo  {journal} {Phys. Rev. B}\ }\textbf {\bibinfo
  {volume} {54}},\ \bibinfo {pages} {7366} (\bibinfo {year}
  {1996})}\BibitemShut {NoStop}%
\bibitem [{\citenamefont {Jonckheere}\ \emph {et~al.}(2013)\citenamefont
  {Jonckheere}, \citenamefont {Rech}, \citenamefont {Martin}, \citenamefont
  {Dou\ifmmode~\mbox{\c{c}}\else \c{c}\fi{}ot}, \citenamefont {Feinberg},\ and\
  \citenamefont {M\'elin}}]{Jonckheere2013}%
  \BibitemOpen
  \bibfield  {author} {\bibinfo {author} {\bibfnamefont {T.}~\bibnamefont
  {Jonckheere}}, \bibinfo {author} {\bibfnamefont {J.}~\bibnamefont {Rech}},
  \bibinfo {author} {\bibfnamefont {T.}~\bibnamefont {Martin}}, \bibinfo
  {author} {\bibfnamefont {B.}~\bibnamefont {Dou\ifmmode~\mbox{\c{c}}\else
  \c{c}\fi{}ot}}, \bibinfo {author} {\bibfnamefont {D.}~\bibnamefont
  {Feinberg}}, \ and\ \bibinfo {author} {\bibfnamefont {R.}~\bibnamefont
  {M\'elin}},\ }\href {\doibase 10.1103/PhysRevB.87.214501} {\bibfield
  {journal} {\bibinfo  {journal} {Phys. Rev. B}\ }\textbf {\bibinfo {volume}
  {87}},\ \bibinfo {pages} {214501} (\bibinfo {year} {2013})}\BibitemShut
  {NoStop}%
\bibitem [{\citenamefont {Martin}(1996)}]{martin96}%
  \BibitemOpen
  \bibfield  {author} {\bibinfo {author} {\bibfnamefont {T.}~\bibnamefont
  {Martin}},\ }\href {\doibase https://doi.org/10.1016/0375-9601(96)00484-7}
  {\bibfield  {journal} {\bibinfo  {journal} {Physics Letters A}\ }\textbf
  {\bibinfo {volume} {220}},\ \bibinfo {pages} {137} (\bibinfo {year}
  {1996})}\BibitemShut {NoStop}%
\bibitem [{\citenamefont {Torr{\`e}s}\ and\ \citenamefont
  {Martin}(1999)}]{torres1999}%
  \BibitemOpen
  \bibfield  {author} {\bibinfo {author} {\bibfnamefont {J.}~\bibnamefont
  {Torr{\`e}s}}\ and\ \bibinfo {author} {\bibfnamefont {T.}~\bibnamefont
  {Martin}},\ }\href {\doibase 10.1007/s100510051010} {\bibfield  {journal}
  {\bibinfo  {journal} {The European Physical Journal B - Condensed Matter and
  Complex Systems}\ }\textbf {\bibinfo {volume} {12}},\ \bibinfo {pages} {319}
  (\bibinfo {year} {1999})}\BibitemShut {NoStop}%
\bibitem [{\citenamefont {Lesovik}\ \emph {et~al.}(2001)\citenamefont
  {Lesovik}, \citenamefont {Martin},\ and\ \citenamefont
  {Blatter}}]{lesovik2001}%
  \BibitemOpen
  \bibfield  {author} {\bibinfo {author} {\bibfnamefont {G.~B.}\ \bibnamefont
  {Lesovik}}, \bibinfo {author} {\bibfnamefont {T.}~\bibnamefont {Martin}}, \
  and\ \bibinfo {author} {\bibfnamefont {G.}~\bibnamefont {Blatter}},\ }\href
  {\doibase 10.1007/s10051-001-8675-4} {\bibfield  {journal} {\bibinfo
  {journal} {The European Physical Journal B - Condensed Matter and Complex
  Systems}\ }\textbf {\bibinfo {volume} {24}},\ \bibinfo {pages} {287}
  (\bibinfo {year} {2001})}\BibitemShut {NoStop}%
\bibitem [{\citenamefont {Recher}\ \emph {et~al.}(2001)\citenamefont {Recher},
  \citenamefont {Sukhorukov},\ and\ \citenamefont {Loss}}]{recher2001}%
  \BibitemOpen
  \bibfield  {author} {\bibinfo {author} {\bibfnamefont {P.}~\bibnamefont
  {Recher}}, \bibinfo {author} {\bibfnamefont {E.~V.}\ \bibnamefont
  {Sukhorukov}}, \ and\ \bibinfo {author} {\bibfnamefont {D.}~\bibnamefont
  {Loss}},\ }\href {\doibase 10.1103/PhysRevB.63.165314} {\bibfield  {journal}
  {\bibinfo  {journal} {Phys. Rev. B}\ }\textbf {\bibinfo {volume} {63}},\
  \bibinfo {pages} {165314} (\bibinfo {year} {2001})}\BibitemShut {NoStop}%
\bibitem [{\citenamefont {Sauret}\ \emph {et~al.}(2004)\citenamefont {Sauret},
  \citenamefont {Feinberg},\ and\ \citenamefont {Martin}}]{sauret2004}%
  \BibitemOpen
  \bibfield  {author} {\bibinfo {author} {\bibfnamefont {O.}~\bibnamefont
  {Sauret}}, \bibinfo {author} {\bibfnamefont {D.}~\bibnamefont {Feinberg}}, \
  and\ \bibinfo {author} {\bibfnamefont {T.}~\bibnamefont {Martin}},\ }\href
  {\doibase 10.1103/PhysRevB.70.245313} {\bibfield  {journal} {\bibinfo
  {journal} {Phys. Rev. B}\ }\textbf {\bibinfo {volume} {70}},\ \bibinfo
  {pages} {245313} (\bibinfo {year} {2004})}\BibitemShut {NoStop}%
\bibitem [{\citenamefont {Chevallier}\ \emph {et~al.}(2011)\citenamefont
  {Chevallier}, \citenamefont {Rech}, \citenamefont {Jonckheere},\ and\
  \citenamefont {Martin}}]{chevallier2011}%
  \BibitemOpen
  \bibfield  {author} {\bibinfo {author} {\bibfnamefont {D.}~\bibnamefont
  {Chevallier}}, \bibinfo {author} {\bibfnamefont {J.}~\bibnamefont {Rech}},
  \bibinfo {author} {\bibfnamefont {T.}~\bibnamefont {Jonckheere}}, \ and\
  \bibinfo {author} {\bibfnamefont {T.}~\bibnamefont {Martin}},\ }\href
  {\doibase 10.1103/PhysRevB.83.125421} {\bibfield  {journal} {\bibinfo
  {journal} {Phys. Rev. B}\ }\textbf {\bibinfo {volume} {83}},\ \bibinfo
  {pages} {125421} (\bibinfo {year} {2011})}\BibitemShut {NoStop}%
\bibitem [{\citenamefont {Rech}\ \emph {et~al.}(2012)\citenamefont {Rech},
  \citenamefont {Chevallier}, \citenamefont {Jonckheere},\ and\ \citenamefont
  {Martin}}]{rech2012}%
  \BibitemOpen
  \bibfield  {author} {\bibinfo {author} {\bibfnamefont {J.}~\bibnamefont
  {Rech}}, \bibinfo {author} {\bibfnamefont {D.}~\bibnamefont {Chevallier}},
  \bibinfo {author} {\bibfnamefont {T.}~\bibnamefont {Jonckheere}}, \ and\
  \bibinfo {author} {\bibfnamefont {T.}~\bibnamefont {Martin}},\ }\href
  {\doibase 10.1103/PhysRevB.85.035419} {\bibfield  {journal} {\bibinfo
  {journal} {Phys. Rev. B}\ }\textbf {\bibinfo {volume} {85}},\ \bibinfo
  {pages} {035419} (\bibinfo {year} {2012})}\BibitemShut {NoStop}%
\bibitem [{\citenamefont {Tsui}(1999)}]{tsui99}%
  \BibitemOpen
  \bibfield  {author} {\bibinfo {author} {\bibfnamefont {D.~C.}\ \bibnamefont
  {Tsui}},\ }\href {\doibase 10.1103/RevModPhys.71.891} {\bibfield  {journal}
  {\bibinfo  {journal} {Rev. Mod. Phys.}\ }\textbf {\bibinfo {volume} {71}},\
  \bibinfo {pages} {891} (\bibinfo {year} {1999})}\BibitemShut {NoStop}%
\bibitem [{\citenamefont {Laughlin}(1983)}]{laughlin83}%
  \BibitemOpen
  \bibfield  {author} {\bibinfo {author} {\bibfnamefont {R.~B.}\ \bibnamefont
  {Laughlin}},\ }\href@noop {} {\bibfield  {journal} {\bibinfo  {journal}
  {Phys. Rev. Lett.}\ }\textbf {\bibinfo {volume} {50}},\ \bibinfo {pages}
  {1395} (\bibinfo {year} {1983})}\BibitemShut {NoStop}%
\bibitem [{\citenamefont {Sandler}\ \emph {et~al.}(1998)\citenamefont
  {Sandler}, \citenamefont {Chamon},\ and\ \citenamefont
  {Fradkin}}]{Sandler1998}%
  \BibitemOpen
  \bibfield  {author} {\bibinfo {author} {\bibfnamefont {N.~P.}\ \bibnamefont
  {Sandler}}, \bibinfo {author} {\bibfnamefont {C.~d.~C.}\ \bibnamefont
  {Chamon}}, \ and\ \bibinfo {author} {\bibfnamefont {E.}~\bibnamefont
  {Fradkin}},\ }\href {\doibase 10.1103/PhysRevB.57.12324} {\bibfield
  {journal} {\bibinfo  {journal} {Phys. Rev. B}\ }\textbf {\bibinfo {volume}
  {57}},\ \bibinfo {pages} {12324} (\bibinfo {year} {1998})}\BibitemShut
  {NoStop}%
\bibitem [{\citenamefont {Hashisaka}\ \emph {et~al.}(2021)\citenamefont
  {Hashisaka}, \citenamefont {Jonckheere}, \citenamefont {Akiho}, \citenamefont
  {Sasaki}, \citenamefont {Rech}, \citenamefont {Martin},\ and\ \citenamefont
  {Muraki}}]{Hashisaka2021}%
  \BibitemOpen
  \bibfield  {author} {\bibinfo {author} {\bibfnamefont {M.}~\bibnamefont
  {Hashisaka}}, \bibinfo {author} {\bibfnamefont {T.}~\bibnamefont
  {Jonckheere}}, \bibinfo {author} {\bibfnamefont {T.}~\bibnamefont {Akiho}},
  \bibinfo {author} {\bibfnamefont {S.}~\bibnamefont {Sasaki}}, \bibinfo
  {author} {\bibfnamefont {J.}~\bibnamefont {Rech}}, \bibinfo {author}
  {\bibfnamefont {T.}~\bibnamefont {Martin}}, \ and\ \bibinfo {author}
  {\bibfnamefont {K.}~\bibnamefont {Muraki}},\ }\href {\doibase
  10.1038/s41467-021-23160-6} {\bibfield  {journal} {\bibinfo  {journal}
  {Nature Communications}\ }\textbf {\bibinfo {volume} {12}},\ \bibinfo {pages}
  {2794} (\bibinfo {year} {2021})}\BibitemShut {NoStop}%
\bibitem [{\citenamefont {Cohen}\ \emph {et~al.}(2022)\citenamefont {Cohen},
  \citenamefont {Samuelson}, \citenamefont {Wang}, \citenamefont {Taniguchi},
  \citenamefont {Watanabe}, \citenamefont {Zaletel},\ and\ \citenamefont
  {Young}}]{cohen2022}%
  \BibitemOpen
  \bibfield  {author} {\bibinfo {author} {\bibfnamefont {L.~A.}\ \bibnamefont
  {Cohen}}, \bibinfo {author} {\bibfnamefont {N.~L.}\ \bibnamefont
  {Samuelson}}, \bibinfo {author} {\bibfnamefont {T.}~\bibnamefont {Wang}},
  \bibinfo {author} {\bibfnamefont {T.}~\bibnamefont {Taniguchi}}, \bibinfo
  {author} {\bibfnamefont {K.}~\bibnamefont {Watanabe}}, \bibinfo {author}
  {\bibfnamefont {M.~P.}\ \bibnamefont {Zaletel}}, \ and\ \bibinfo {author}
  {\bibfnamefont {A.~F.}\ \bibnamefont {Young}},\ }\href@noop {} {\enquote
  {\bibinfo {title} {Universal chiral luttinger liquid behavior in a graphene
  fractional quantum hall point contact},}\ } (\bibinfo {year} {2022}),\
  \Eprint {http://arxiv.org/abs/2212.01374} {arXiv:2212.01374
  [cond-mat.mes-hall]} \BibitemShut {NoStop}%
\bibitem [{\citenamefont {Fukuzawa}\ \emph {et~al.}(2023)\citenamefont
  {Fukuzawa}, \citenamefont {Kato}, \citenamefont {Jonckheere}, \citenamefont
  {Rech},\ and\ \citenamefont {Martin}}]{fukuzawa2023minimal}%
  \BibitemOpen
  \bibfield  {author} {\bibinfo {author} {\bibfnamefont {K.}~\bibnamefont
  {Fukuzawa}}, \bibinfo {author} {\bibfnamefont {T.}~\bibnamefont {Kato}},
  \bibinfo {author} {\bibfnamefont {T.}~\bibnamefont {Jonckheere}}, \bibinfo
  {author} {\bibfnamefont {J.}~\bibnamefont {Rech}}, \ and\ \bibinfo {author}
  {\bibfnamefont {T.}~\bibnamefont {Martin}},\ }\href@noop {} {\enquote
  {\bibinfo {title} {Minimal ac injection into carbon nanotubes},}\ } (\bibinfo
  {year} {2023}),\ \Eprint {http://arxiv.org/abs/2307.11943} {arXiv:2307.11943
  [cond-mat.mes-hall]} \BibitemShut {NoStop}%
\bibitem [{\citenamefont {Kane}\ and\ \citenamefont {Fisher}(2003)}]{Kane2003}%
  \BibitemOpen
  \bibfield  {author} {\bibinfo {author} {\bibfnamefont {C.~L.}\ \bibnamefont
  {Kane}}\ and\ \bibinfo {author} {\bibfnamefont {M.~P.~A.}\ \bibnamefont
  {Fisher}},\ }\href {\doibase 10.1103/PhysRevB.67.045307} {\bibfield
  {journal} {\bibinfo  {journal} {Phys. Rev. B}\ }\textbf {\bibinfo {volume}
  {67}},\ \bibinfo {pages} {045307} (\bibinfo {year} {2003})}\BibitemShut
  {NoStop}%
\bibitem [{\citenamefont {Comforti}\ \emph {et~al.}(2002)\citenamefont
  {Comforti}, \citenamefont {Chung}, \citenamefont {Heiblum}, \citenamefont
  {Umansky},\ and\ \citenamefont {Mahalu}}]{Comforti2002}%
  \BibitemOpen
  \bibfield  {author} {\bibinfo {author} {\bibfnamefont {E.}~\bibnamefont
  {Comforti}}, \bibinfo {author} {\bibfnamefont {Y.~C.}\ \bibnamefont {Chung}},
  \bibinfo {author} {\bibfnamefont {M.}~\bibnamefont {Heiblum}}, \bibinfo
  {author} {\bibfnamefont {V.}~\bibnamefont {Umansky}}, \ and\ \bibinfo
  {author} {\bibfnamefont {D.}~\bibnamefont {Mahalu}},\ }\href {\doibase
  10.1038/416515a} {\bibfield  {journal} {\bibinfo  {journal} {Nature}\
  }\textbf {\bibinfo {volume} {416}},\ \bibinfo {pages} {515} (\bibinfo {year}
  {2002})}\BibitemShut {NoStop}%
\bibitem [{\citenamefont {Chung}\ \emph {et~al.}(2003)\citenamefont {Chung},
  \citenamefont {Heiblum}, \citenamefont {Oreg}, \citenamefont {Umansky},\ and\
  \citenamefont {Mahalu}}]{Chung2003}%
  \BibitemOpen
  \bibfield  {author} {\bibinfo {author} {\bibfnamefont {Y.~C.}\ \bibnamefont
  {Chung}}, \bibinfo {author} {\bibfnamefont {M.}~\bibnamefont {Heiblum}},
  \bibinfo {author} {\bibfnamefont {Y.}~\bibnamefont {Oreg}}, \bibinfo {author}
  {\bibfnamefont {V.}~\bibnamefont {Umansky}}, \ and\ \bibinfo {author}
  {\bibfnamefont {D.}~\bibnamefont {Mahalu}},\ }\href {\doibase
  10.1103/PhysRevB.67.201104} {\bibfield  {journal} {\bibinfo  {journal} {Phys.
  Rev. B}\ }\textbf {\bibinfo {volume} {67}},\ \bibinfo {pages} {201104}
  (\bibinfo {year} {2003})}\BibitemShut {NoStop}%
\bibitem [{\citenamefont {Glidic}\ \emph
  {et~al.}(2023{\natexlab{a}})\citenamefont {Glidic}, \citenamefont {Maillet},
  \citenamefont {Piquard}, \citenamefont {Aassime}, \citenamefont {Cavanna},
  \citenamefont {Jin}, \citenamefont {Gennser}, \citenamefont {Anthore},\ and\
  \citenamefont {Pierre}}]{Glidic2023}%
  \BibitemOpen
  \bibfield  {author} {\bibinfo {author} {\bibfnamefont {P.}~\bibnamefont
  {Glidic}}, \bibinfo {author} {\bibfnamefont {O.}~\bibnamefont {Maillet}},
  \bibinfo {author} {\bibfnamefont {C.}~\bibnamefont {Piquard}}, \bibinfo
  {author} {\bibfnamefont {A.}~\bibnamefont {Aassime}}, \bibinfo {author}
  {\bibfnamefont {A.}~\bibnamefont {Cavanna}}, \bibinfo {author} {\bibfnamefont
  {Y.}~\bibnamefont {Jin}}, \bibinfo {author} {\bibfnamefont {U.}~\bibnamefont
  {Gennser}}, \bibinfo {author} {\bibfnamefont {A.}~\bibnamefont {Anthore}}, \
  and\ \bibinfo {author} {\bibfnamefont {F.}~\bibnamefont {Pierre}},\ }\href
  {\doibase 10.1038/s41467-023-36080-4} {\bibfield  {journal} {\bibinfo
  {journal} {Nature Communications}\ }\textbf {\bibinfo {volume} {14}},\
  \bibinfo {pages} {514} (\bibinfo {year} {2023}{\natexlab{a}})}\BibitemShut
  {NoStop}%
\bibitem [{\citenamefont {Bartolomei}\ \emph {et~al.}(2020)\citenamefont
  {Bartolomei}, \citenamefont {Kumar}, \citenamefont {Bisognin}, \citenamefont
  {Marguerite}, \citenamefont {Berroir}, \citenamefont {Bocquillon},
  \citenamefont {Plaçais}, \citenamefont {Cavanna}, \citenamefont {Dong},
  \citenamefont {Gennser},\ and\ \citenamefont {et~al.}}]{bartolomei20}%
  \BibitemOpen
  \bibfield  {author} {\bibinfo {author} {\bibfnamefont {H.}~\bibnamefont
  {Bartolomei}}, \bibinfo {author} {\bibfnamefont {M.}~\bibnamefont {Kumar}},
  \bibinfo {author} {\bibfnamefont {R.}~\bibnamefont {Bisognin}}, \bibinfo
  {author} {\bibfnamefont {A.}~\bibnamefont {Marguerite}}, \bibinfo {author}
  {\bibfnamefont {J.-M.}\ \bibnamefont {Berroir}}, \bibinfo {author}
  {\bibfnamefont {E.}~\bibnamefont {Bocquillon}}, \bibinfo {author}
  {\bibfnamefont {B.}~\bibnamefont {Plaçais}}, \bibinfo {author}
  {\bibfnamefont {A.}~\bibnamefont {Cavanna}}, \bibinfo {author} {\bibfnamefont
  {Q.}~\bibnamefont {Dong}}, \bibinfo {author} {\bibfnamefont {U.}~\bibnamefont
  {Gennser}}, \ and\ \bibinfo {author} {\bibnamefont {et~al.}},\ }\href
  {\doibase 10.1126/science.aaz5601} {\bibfield  {journal} {\bibinfo  {journal}
  {Science}\ }\textbf {\bibinfo {volume} {368}},\ \bibinfo {pages} {173–177}
  (\bibinfo {year} {2020})}\BibitemShut {NoStop}%
\bibitem [{\citenamefont {Ruelle}\ \emph {et~al.}(2023)\citenamefont {Ruelle},
  \citenamefont {Frigerio}, \citenamefont {Berroir}, \citenamefont
  {Pla\ifmmode~\mbox{\c{c}}\else \c{c}\fi{}ais}, \citenamefont {Rech},
  \citenamefont {Cavanna}, \citenamefont {Gennser}, \citenamefont {Jin},\ and\
  \citenamefont {F\`eve}}]{Ruelle2022}%
  \BibitemOpen
  \bibfield  {author} {\bibinfo {author} {\bibfnamefont {M.}~\bibnamefont
  {Ruelle}}, \bibinfo {author} {\bibfnamefont {E.}~\bibnamefont {Frigerio}},
  \bibinfo {author} {\bibfnamefont {J.-M.}\ \bibnamefont {Berroir}}, \bibinfo
  {author} {\bibfnamefont {B.}~\bibnamefont {Pla\ifmmode~\mbox{\c{c}}\else
  \c{c}\fi{}ais}}, \bibinfo {author} {\bibfnamefont {J.}~\bibnamefont {Rech}},
  \bibinfo {author} {\bibfnamefont {A.}~\bibnamefont {Cavanna}}, \bibinfo
  {author} {\bibfnamefont {U.}~\bibnamefont {Gennser}}, \bibinfo {author}
  {\bibfnamefont {Y.}~\bibnamefont {Jin}}, \ and\ \bibinfo {author}
  {\bibfnamefont {G.}~\bibnamefont {F\`eve}},\ }\href {\doibase
  10.1103/PhysRevX.13.011031} {\bibfield  {journal} {\bibinfo  {journal} {Phys.
  Rev. X}\ }\textbf {\bibinfo {volume} {13}},\ \bibinfo {pages} {011031}
  (\bibinfo {year} {2023})}\BibitemShut {NoStop}%
\bibitem [{\citenamefont {Glidic}\ \emph
  {et~al.}(2023{\natexlab{b}})\citenamefont {Glidic}, \citenamefont {Maillet},
  \citenamefont {Aassime}, \citenamefont {Piquard}, \citenamefont {Cavanna},
  \citenamefont {Gennser}, \citenamefont {Jin}, \citenamefont {Anthore},\ and\
  \citenamefont {Pierre}}]{Glidic2022braiding}%
  \BibitemOpen
  \bibfield  {author} {\bibinfo {author} {\bibfnamefont {P.}~\bibnamefont
  {Glidic}}, \bibinfo {author} {\bibfnamefont {O.}~\bibnamefont {Maillet}},
  \bibinfo {author} {\bibfnamefont {A.}~\bibnamefont {Aassime}}, \bibinfo
  {author} {\bibfnamefont {C.}~\bibnamefont {Piquard}}, \bibinfo {author}
  {\bibfnamefont {A.}~\bibnamefont {Cavanna}}, \bibinfo {author} {\bibfnamefont
  {U.}~\bibnamefont {Gennser}}, \bibinfo {author} {\bibfnamefont
  {Y.}~\bibnamefont {Jin}}, \bibinfo {author} {\bibfnamefont {A.}~\bibnamefont
  {Anthore}}, \ and\ \bibinfo {author} {\bibfnamefont {F.}~\bibnamefont
  {Pierre}},\ }\href {\doibase 10.1103/PhysRevX.13.011030} {\bibfield
  {journal} {\bibinfo  {journal} {Phys. Rev. X}\ }\textbf {\bibinfo {volume}
  {13}},\ \bibinfo {pages} {011030} (\bibinfo {year}
  {2023}{\natexlab{b}})}\BibitemShut {NoStop}%
\bibitem [{\citenamefont {Lee}\ \emph {et~al.}(2023)\citenamefont {Lee},
  \citenamefont {Hong}, \citenamefont {Alkalay}, \citenamefont {Schiller},
  \citenamefont {Umansky}, \citenamefont {Heiblum}, \citenamefont {Oreg},\ and\
  \citenamefont {Sim}}]{Lee2022}%
  \BibitemOpen
  \bibfield  {author} {\bibinfo {author} {\bibfnamefont {J.-Y.~M.}\
  \bibnamefont {Lee}}, \bibinfo {author} {\bibfnamefont {C.}~\bibnamefont
  {Hong}}, \bibinfo {author} {\bibfnamefont {T.}~\bibnamefont {Alkalay}},
  \bibinfo {author} {\bibfnamefont {N.}~\bibnamefont {Schiller}}, \bibinfo
  {author} {\bibfnamefont {V.}~\bibnamefont {Umansky}}, \bibinfo {author}
  {\bibfnamefont {M.}~\bibnamefont {Heiblum}}, \bibinfo {author} {\bibfnamefont
  {Y.}~\bibnamefont {Oreg}}, \ and\ \bibinfo {author} {\bibfnamefont {H.-S.}\
  \bibnamefont {Sim}},\ }\href {\doibase 10.1038/s41586-023-05883-2} {\bibfield
   {journal} {\bibinfo  {journal} {Nature}\ }\textbf {\bibinfo {volume}
  {617}},\ \bibinfo {pages} {277} (\bibinfo {year} {2023})}\BibitemShut
  {NoStop}%
\bibitem [{\citenamefont {Safi}\ \emph {et~al.}(2001)\citenamefont {Safi},
  \citenamefont {Devillard},\ and\ \citenamefont {Martin}}]{safi01}%
  \BibitemOpen
  \bibfield  {author} {\bibinfo {author} {\bibfnamefont {I.}~\bibnamefont
  {Safi}}, \bibinfo {author} {\bibfnamefont {P.}~\bibnamefont {Devillard}}, \
  and\ \bibinfo {author} {\bibfnamefont {T.}~\bibnamefont {Martin}},\ }\href
  {\doibase 10.1103/PhysRevLett.86.4628} {\bibfield  {journal} {\bibinfo
  {journal} {Phys. Rev. Lett.}\ }\textbf {\bibinfo {volume} {86}},\ \bibinfo
  {pages} {4628} (\bibinfo {year} {2001})}\BibitemShut {NoStop}%
\bibitem [{\citenamefont {Kane}(2003)}]{kane03}%
  \BibitemOpen
  \bibfield  {author} {\bibinfo {author} {\bibfnamefont {C.~L.}\ \bibnamefont
  {Kane}},\ }\href {\doibase 10.1103/PhysRevLett.90.226802} {\bibfield
  {journal} {\bibinfo  {journal} {Phys. Rev. Lett.}\ }\textbf {\bibinfo
  {volume} {90}},\ \bibinfo {pages} {226802} (\bibinfo {year}
  {2003})}\BibitemShut {NoStop}%
\bibitem [{\citenamefont {Jonckheere}\ \emph {et~al.}(2005)\citenamefont
  {Jonckheere}, \citenamefont {Creux},\ and\ \citenamefont
  {Martin}}]{jonckheere05}%
  \BibitemOpen
  \bibfield  {author} {\bibinfo {author} {\bibfnamefont {T.}~\bibnamefont
  {Jonckheere}}, \bibinfo {author} {\bibfnamefont {M.}~\bibnamefont {Creux}}, \
  and\ \bibinfo {author} {\bibfnamefont {T.}~\bibnamefont {Martin}},\
  }\href@noop {} {\bibfield  {journal} {\bibinfo  {journal} {Phys. Rev. B}\
  }\textbf {\bibinfo {volume} {72}},\ \bibinfo {pages} {205321} (\bibinfo
  {year} {2005})}\BibitemShut {NoStop}%
\bibitem [{\citenamefont {Lee}\ \emph {et~al.}(2019)\citenamefont {Lee},
  \citenamefont {Han},\ and\ \citenamefont {Sim}}]{Lee19}%
  \BibitemOpen
  \bibfield  {author} {\bibinfo {author} {\bibfnamefont {B.}~\bibnamefont
  {Lee}}, \bibinfo {author} {\bibfnamefont {C.}~\bibnamefont {Han}}, \ and\
  \bibinfo {author} {\bibfnamefont {H.-S.}\ \bibnamefont {Sim}},\ }\href
  {\doibase 10.1103/PhysRevLett.123.016803} {\bibfield  {journal} {\bibinfo
  {journal} {Phys. Rev. Lett.}\ }\textbf {\bibinfo {volume} {123}},\ \bibinfo
  {pages} {016803} (\bibinfo {year} {2019})}\BibitemShut {NoStop}%
\bibitem [{\citenamefont {Nayak}\ \emph {et~al.}(2008)\citenamefont {Nayak},
  \citenamefont {Simon}, \citenamefont {Stern}, \citenamefont {Freedman},\ and\
  \citenamefont {Sarma}}]{nayak08}%
  \BibitemOpen
  \bibfield  {author} {\bibinfo {author} {\bibfnamefont {C.}~\bibnamefont
  {Nayak}}, \bibinfo {author} {\bibfnamefont {S.~H.}\ \bibnamefont {Simon}},
  \bibinfo {author} {\bibfnamefont {A.}~\bibnamefont {Stern}}, \bibinfo
  {author} {\bibfnamefont {M.}~\bibnamefont {Freedman}}, \ and\ \bibinfo
  {author} {\bibfnamefont {S.~D.}\ \bibnamefont {Sarma}},\ }\href@noop {}
  {\bibfield  {journal} {\bibinfo  {journal} {Rev. Mod. Phys.}\ }\textbf
  {\bibinfo {volume} {80}},\ \bibinfo {pages} {1083} (\bibinfo {year}
  {2008})}\BibitemShut {NoStop}%
\bibitem [{\citenamefont {Levkivskyi}(2016)}]{Levkivskyi16}%
  \BibitemOpen
  \bibfield  {author} {\bibinfo {author} {\bibfnamefont {I.~P.}\ \bibnamefont
  {Levkivskyi}},\ }\href {\doibase 10.1103/PhysRevB.93.165427} {\bibfield
  {journal} {\bibinfo  {journal} {Phys. Rev. B}\ }\textbf {\bibinfo {volume}
  {93}},\ \bibinfo {pages} {165427} (\bibinfo {year} {2016})}\BibitemShut
  {NoStop}%
\bibitem [{\citenamefont {Rosenow}\ \emph {et~al.}(2016)\citenamefont
  {Rosenow}, \citenamefont {Levkivskyi},\ and\ \citenamefont
  {Halperin}}]{rosenow16}%
  \BibitemOpen
  \bibfield  {author} {\bibinfo {author} {\bibfnamefont {B.}~\bibnamefont
  {Rosenow}}, \bibinfo {author} {\bibfnamefont {I.~P.}\ \bibnamefont
  {Levkivskyi}}, \ and\ \bibinfo {author} {\bibfnamefont {B.~I.}\ \bibnamefont
  {Halperin}},\ }\href {\doibase 10.1103/PhysRevLett.116.156802} {\bibfield
  {journal} {\bibinfo  {journal} {Phys. Rev. Lett.}\ }\textbf {\bibinfo
  {volume} {116}},\ \bibinfo {pages} {156802} (\bibinfo {year}
  {2016})}\BibitemShut {NoStop}%
\bibitem [{\citenamefont {Martin}(2005)}]{martin05}%
  \BibitemOpen
  \bibfield  {author} {\bibinfo {author} {\bibfnamefont {T.}~\bibnamefont
  {Martin}},\ }in\ \href@noop {} {\emph {\bibinfo {booktitle} {Nanophysics:
  Coherence and Transport}}},\ \bibinfo {series and number} {Les Houches,
  Session LXXXI},\ \bibinfo {editor} {edited by\ \bibinfo {editor}
  {\bibfnamefont {H.}~\bibnamefont {Bouchiat}}, \bibinfo {editor}
  {\bibfnamefont {Y.}~\bibnamefont {Gefen}}, \bibinfo {editor} {\bibfnamefont
  {S.}~\bibnamefont {Gu{\'e}ron}}, \bibinfo {editor} {\bibfnamefont
  {G.}~\bibnamefont {Montambaux}}, \ and\ \bibinfo {editor} {\bibfnamefont
  {J.}~\bibnamefont {Dalibard}}}\ (\bibinfo  {publisher} {Elsevier},\ \bibinfo
  {year} {2005})\ p.\ \bibinfo {pages} {283}\BibitemShut {NoStop}%
\bibitem [{\citenamefont {Han}\ \emph {et~al.}(2016)\citenamefont {Han},
  \citenamefont {Park}, \citenamefont {Gefen},\ and\ \citenamefont
  {Sim}}]{Han2016}%
  \BibitemOpen
  \bibfield  {author} {\bibinfo {author} {\bibfnamefont {C.}~\bibnamefont
  {Han}}, \bibinfo {author} {\bibfnamefont {J.}~\bibnamefont {Park}}, \bibinfo
  {author} {\bibfnamefont {Y.}~\bibnamefont {Gefen}}, \ and\ \bibinfo {author}
  {\bibfnamefont {H.-S.}\ \bibnamefont {Sim}},\ }\href {\doibase
  10.1038/ncomms11131} {\bibfield  {journal} {\bibinfo  {journal} {Nature
  Communications}\ }\textbf {\bibinfo {volume} {7}},\ \bibinfo {pages} {11131}
  (\bibinfo {year} {2016})}\BibitemShut {NoStop}%
\bibitem [{\citenamefont {Morel}\ \emph {et~al.}(2022)\citenamefont {Morel},
  \citenamefont {Lee}, \citenamefont {Sim},\ and\ \citenamefont
  {Mora}}]{Morel22}%
  \BibitemOpen
  \bibfield  {author} {\bibinfo {author} {\bibfnamefont {T.}~\bibnamefont
  {Morel}}, \bibinfo {author} {\bibfnamefont {J.-Y.~M.}\ \bibnamefont {Lee}},
  \bibinfo {author} {\bibfnamefont {H.-S.}\ \bibnamefont {Sim}}, \ and\
  \bibinfo {author} {\bibfnamefont {C.}~\bibnamefont {Mora}},\ }\href {\doibase
  10.1103/PhysRevB.105.075433} {\bibfield  {journal} {\bibinfo  {journal}
  {Phys. Rev. B}\ }\textbf {\bibinfo {volume} {105}},\ \bibinfo {pages}
  {075433} (\bibinfo {year} {2022})}\BibitemShut {NoStop}%
\bibitem [{\citenamefont {Wen}(1995)}]{wen95}%
  \BibitemOpen
  \bibfield  {author} {\bibinfo {author} {\bibfnamefont {X.~G.}\ \bibnamefont
  {Wen}},\ }\href@noop {} {\bibfield  {journal} {\bibinfo  {journal} {Adv.
  Phys.}\ }\textbf {\bibinfo {volume} {44}},\ \bibinfo {pages} {405} (\bibinfo
  {year} {1995})}\BibitemShut {NoStop}%
\bibitem [{\citenamefont {Gradshteyn}\ \emph {et~al.}(2014)\citenamefont
  {Gradshteyn}, \citenamefont {Ryzhik}, \citenamefont {Zwillinger},\ and\
  \citenamefont {Moll}}]{gradshteyn14}%
  \BibitemOpen
  \bibfield  {author} {\bibinfo {author} {\bibfnamefont {I.~S.}\ \bibnamefont
  {Gradshteyn}}, \bibinfo {author} {\bibfnamefont {I.~M.}\ \bibnamefont
  {Ryzhik}}, \bibinfo {author} {\bibfnamefont {D.}~\bibnamefont {Zwillinger}},
  \ and\ \bibinfo {author} {\bibfnamefont {V.}~\bibnamefont {Moll}},\
  }\href@noop {} {\emph {\bibinfo {title} {{Table of integrals, series, and
  products; 8th ed.}}}}\ (\bibinfo  {publisher} {Academic Press},\ \bibinfo
  {address} {Amsterdam},\ \bibinfo {year} {2014})\BibitemShut {NoStop}%
\bibitem [{\citenamefont {Shtanko}\ \emph {et~al.}(2014)\citenamefont
  {Shtanko}, \citenamefont {Snizhko},\ and\ \citenamefont
  {Cheianov}}]{shtanko14}%
  \BibitemOpen
  \bibfield  {author} {\bibinfo {author} {\bibfnamefont {O.}~\bibnamefont
  {Shtanko}}, \bibinfo {author} {\bibfnamefont {K.}~\bibnamefont {Snizhko}}, \
  and\ \bibinfo {author} {\bibfnamefont {V.}~\bibnamefont {Cheianov}},\ }\href
  {\doibase 10.1103/PhysRevB.89.125104} {\bibfield  {journal} {\bibinfo
  {journal} {Phys. Rev. B}\ }\textbf {\bibinfo {volume} {89}},\ \bibinfo
  {pages} {125104} (\bibinfo {year} {2014})}\BibitemShut {NoStop}%
\end{thebibliography}%

%
%

\onecolumngrid
\appendix

\section{Useful contour integrals}
\label{app:contour}

First, we need to evaluate the following integral
\begin{align}
 J_1^{\eta_3 \eta_4}(t_2,\tau) &=  \int dT  \frac{1}{\left[T -  \frac{\tau}{2} + i\eta_4 \tau_0 \right]\left[T - t_2 + \frac{\tau}{2} + i \eta_3 \tau_0\right]} .
\end{align}
We will use the residue theorem to go about this. $J_1$ has two single poles located at $T = \frac{\tau}{2} - i\eta_4 \tau_0$ and $T = t_2 - \frac{\tau}{2} - i\eta_3 \tau_0$. Firstly, we note that if $\eta_3 = \eta_4$, then both poles lie on the same side of the real axis. This means that we can close the contour over the opposite half-plane which encloses no poles and hence gives zero, that is, $J_1^{\eta \eta}(t_2,\tau) = 0$. 

We then focus on the only nonzero case, namely $\eta_4 = -\eta_3$.
We choose to always close the contour over the half-plane enclosing the pole at $T = \frac{\tau}{2} -i\eta_4 \tau_0 $, thus picking an overall sign $\eta_3$, ultimately yielding
\begin{equation}
 J_1^{\eta_3 \eta_4}(t_2,\tau) = \frac{2i \pi \eta_3 }{\tau - t_2  + 2 i \eta_3 \tau_0} \delta_{\eta_3,-\eta_4} .
  \label{Tintegral}
\end{equation}

The second integral of interest reads
\begin{equation}
J_2^{\eta_1 \eta_3}(\tau)  = \int dt_2 \frac{t_2}{\left(t_2 - i \eta_1 \tau_0 \right)^\frac{2}{\nu}(t_2-\tau -2i\eta_3 \tau_0)} .
\end{equation}
The integrand of $J_2$ has a relatively complicated $\frac{2}{\nu}$-order pole at $t_2 = i \eta_1 \tau_0$, and a simple first-order pole at $t_2 = \tau + 2i\eta_3 \tau_0$. As for $J_1$, the resulting integral vanished if both poles lie on the same half-plane, i.e. if $\eta_1 = \eta_3$. We thus focus again on the only nonzero case, namely $\eta_1 = - \eta_3$. We choose to close the contour over the half-plane enclosing the single pole, thus picking up an overall sign $\eta_3$, ultimately yielding
\begin{equation}
J_2^{\eta_1 \eta_3}(\tau)  = 2 i \pi \eta_3 \frac{\tau + 2i \eta_3 \tau_0}{\left( \tau + 3 i \eta_3 \tau_0\right)^{2/\nu}} \delta_{\eta_3,-\eta_1}  .
\end{equation}

\section{Finite temperature calculation}
\label{app:finiteT}

Several elements of the derivation are common between auto- and cross-correlations and we try to present the two together as much as possible.

Our starting point is given by the auto- and cross-correlations expressed in terms of the bosonic Green's function, and obtained from Eqs.~\eqref{cross_corr_intermediate} and \eqref{eq:auto_corr_intermediate}
\begin{align}
S_{23} &= \frac{2e^2 v_F^2}{(2\pi i)^2}\frac{\Gamma_L^2\Gamma_R^2}{(2\pi a)^{2\nu + 2/
    \nu}} \int dt dt_1 \dots dt_4 \sum_{\{ \eta_k \}} \eta_1 \dots \eta_4 ~ e^{ie^* V t_{34}} e^{2\nu\mathcal{G}^{\eta_{3}\eta_{4}}(0, t_{34})}  e^{\frac{2}{\nu}\mathcal{G}^{\eta_{1}\eta_{2}}(0, t_{12})} \nonumber  \\  
& \times \partial_x \left[ - \mathcal{G}^{-\eta_1}(x-L,t-t_1) + \mathcal{G}^{-\eta_2}(x-L,t-t_2) - \nu\mathcal{G}^{-\eta_3}(x,t-t_3) + \nu\mathcal{G}^{-\eta_4}(x,t-t_4) \right] \nonumber \\
& \times \partial_y \left[ \mathcal{G}^{+\eta_1}(y-L,-t_1) - \mathcal{G}^{+\eta_2}(y-L,-t_2) \right]
\left[ e^{-\mathcal{G}^{\eta_{1}\eta_{3}}(L,t_{13})}e^{\mathcal{G}^{\eta_{1}\eta_{4}}(L,t_{14})}e^{\mathcal{G}^{\eta_{2}\eta_{3}}(L,t_{23})}e^{-\mathcal{G}^{\eta_{2}\eta_{4}}(L,t_{24})}  - 1\right] , \\
S_{33}^\text{exc} &= \frac{2e^2 v_F^2}{(2\pi i)^2}\frac{\Gamma_L^2\Gamma_R^2}{(2\pi a)^{2\nu + 2/
    \nu}} \int dt dt_1 \dots dt_4 \sum_{\{ \eta_k \}} \eta_1 \dots \eta_4 ~ e^{ie^* V t_{34}} e^{2\nu\mathcal{G}^{\eta_{3}\eta_{4}}(0, t_{34})}  e^{\frac{2}{\nu}\mathcal{G}^{\eta_{1}\eta_{2}}(0, t_{12})} \nonumber  \\  
& \times \partial_x \left[ \mathcal{G}^{-\eta_1}(x-L,t-t_1) - \mathcal{G}^{-\eta_2}(x-L,t-t_2) \right] \partial_y \left[ \mathcal{G}^{+\eta_1}(y-L,-t_1) - \mathcal{G}^{+\eta_2}(y-L,-t_2) \right] \nonumber \\
& \times 
\left[ e^{-\mathcal{G}^{\eta_{1}\eta_{3}}(L,t_{13})}e^{\mathcal{G}^{\eta_{1}\eta_{4}}(L,t_{14})}e^{\mathcal{G}^{\eta_{2}\eta_{3}}(L,t_{23})}e^{-\mathcal{G}^{\eta_{2}\eta_{4}}(L,t_{24})}  - 1\right] .
\end{align}

\subsection{$t-$ and $t_1-$integrals}

The first complication arising at finite temperature comes from the the $t-$ and $t_1-$integrals as the Green's function derivative no longer reduces to a simple delta function. Instead of Eq.~\eqref{green_fn_derivative}, one has at finite temperature
\begin{align}
    \partial_x \mathcal{G}^{\eta_1 \eta_2} (x,\tau) = - \frac{\pi \chi_{12} (\tau) \theta/v_F}{\tanh \left[ \pi \theta \left( i \tau_0 - \chi_{12} (\tau) (\tau-x/v_F) \right) \right]}  ,
\end{align}
allowing us to write the general integrated form as (assuming here $x>0$)
\begin{align}
\int dt \partial_x \left[ \mathcal{G}^{\eta \eta_1} (x, t-t_1) -  \mathcal{G}^{\eta \eta_2} (x, t-t_2) \right] = i \frac{\pi (1-2\theta \tau_0)}{v_F} \left( \eta_1 - \eta_2 \right) - \frac{2 \pi \theta}{v_F} \left( t_1 - t_2 \right)    .
\end{align}
This, in turn, enables us to write (up to leading order in the cutoff)
\begin{align}
S_{23} &=\frac{e^2\Gamma_L^2\Gamma_R^2}{2(2\pi a)^{2\nu + 2/
    \nu}} \int  dt_2 dT d\tau \sum_{\{ \eta_k \}} \eta_1 \dots \eta_4 ~ e^{ie^* V \tau} e^{2\nu\mathcal{G}^{\eta_{3}\eta_{4}}(0, \tau)}  e^{\frac{2}{\nu}\mathcal{G}^{\eta_{1}\eta_{2}}(0, t_{2})} 
     \left[ i  \left( \eta_1 - \eta_2 \right) - 2  \theta t_2 \right] \nonumber  \\  
& \times \left[  i  \left( \eta_1 - \eta_2 \right) - 2  \theta t_2 + i  \nu \left( \eta_3 - \eta_4 \right) - 2  \nu \theta \tau \right] 
\left[ \frac{e^{\mathcal{G}^{\eta_{1}\eta_{4}}(L,-T+\tau/2)}e^{\mathcal{G}^{\eta_{2}\eta_{3}}(L,-t_{2}-T-\tau/2)}}{e^{\mathcal{G}^{\eta_{1}\eta_{3}}(L,-T-\tau/2)}e^{\mathcal{G}^{\eta_{2}\eta_{4}}(L,-t_{2}-T+\tau/2)}}   - 1\right]  , \\
S_{33}^\text{exc} &= -\frac{e^2 \Gamma_L^2\Gamma_R^2}{2 (2\pi a)^{2\nu + 2/
    \nu}} \int dt_2 dT d\tau \sum_{\{ \eta_k \}} \eta_1 \dots \eta_4 ~ e^{ie^* V \tau} e^{2\nu\mathcal{G}^{\eta_{3}\eta_{4}}(0, \tau)}  e^{\frac{2}{\nu}\mathcal{G}^{\eta_{1}\eta_{2}}(0, t_{2})} 
    \left[ i  \left( \eta_1 - \eta_2 \right) - 2 \theta t_2 \right]^2  \nonumber  \\ 
& \times 
\left[ \frac{e^{\mathcal{G}^{\eta_{1}\eta_{4}}(L,-T+\tau/2)}e^{\mathcal{G}^{\eta_{2}\eta_{3}}(L,-t_{2}-T-\tau/2)}}{e^{\mathcal{G}^{\eta_{1}\eta_{3}}(L,-T-\tau/2)}e^{\mathcal{G}^{\eta_{2}\eta_{4}}(L,-t_{2}-T+\tau/2)}}   - 1\right]   ,
\end{align}
where we flipped the sign of $t_2$ and, following the same steps as the zero-temperature calculation, we changed variables from $t_3, t_4$ to $T=(t_3+t_4)/2$ and $\tau=t_3-t_4$.

\subsection{$T-$integral}

At this point, one notices that the variable $T$ is only present in the final term, allowing us to perform the corresponding integral yielding
\begin{align}
    F_{\eta_3,\eta_4} (t_2, \tau) &=  \int dT  \left[ \frac{e^{\mathcal{G}^{\eta_{1}\eta_{4}}(L,-T+\tau/2)}e^{\mathcal{G}^{\eta_{2}\eta_{3}}(L,-t_{2}-T-\tau/2)}}{e^{\mathcal{G}^{\eta_{1}\eta_{3}}(L,-T-\tau/2)}e^{\mathcal{G}^{\eta_{2}\eta_{4}}(L,-t_{2}-T+\tau/2)}}   - 1\right] \nonumber \\
    &=  - \frac{2}{\pi \theta} \frac{\sinh \left(\pi \theta t_2 \right) \sinh \left( \pi \theta \tau+i \alpha (\eta_3-\eta_4) \right)}{\sinh \left( \pi \theta (t_2+\tau) + i \alpha (\eta_3 - \eta_4 ) \right)} \left[ \pi \theta \left( t_2+\tau \right) + i \alpha \left( \eta_3 - \eta_4 \right) - i \frac{\pi}{2} \left( \eta_3 - \eta_4 \right) \right]  ,
\end{align}
with the shorthand notation $\alpha = \frac{\pi \theta a}{v_F}$.

This then leaves us with the following expressions for the current correlations
\begin{align}
S_{23} &=\frac{e^2\Gamma_L^2\Gamma_R^2}{2(2\pi a)^{2\nu + 2/
    \nu}} \int  dt_2  d\tau \sum_{\{ \eta_k \}} \eta_1 \dots \eta_4 ~ e^{ie^* V \tau} e^{2\nu\mathcal{G}^{\eta_{3}\eta_{4}}(0, \tau)}  e^{\frac{2}{\nu}\mathcal{G}^{\eta_{1}\eta_{2}}(0, t_{2})} 
     \left[ i  \left( \eta_1 - \eta_2 \right) - 2  \theta t_2 \right] \nonumber  \\  
& \times \left[  i  \left( \eta_1 - \eta_2 \right) - 2  \theta t_2 + i  \nu \left( \eta_3 - \eta_4 \right) - 2  \nu \theta \tau \right] F_{\eta_3,\eta_4} (t_2, \tau)  ,  \\
S_{33}^\text{exc} &= -\frac{e^2 \Gamma_L^2\Gamma_R^2}{2 (2\pi a)^{2\nu + 2/
    \nu}} \int dt_2  d\tau \sum_{\{ \eta_k \}} \eta_1 \dots \eta_4 ~ e^{ie^* V \tau} e^{2\nu\mathcal{G}^{\eta_{3}\eta_{4}}(0, \tau)}  e^{\frac{2}{\nu}\mathcal{G}^{\eta_{1}\eta_{2}}(0, t_{2})} 
    \left[ i  \left( \eta_1 - \eta_2 \right) - 2 \theta t_2 \right]^2  F_{\eta_3,\eta_4} (t_2, \tau)  .
\end{align}

\subsection{$t_2-$integral}

Performing explicitly the summations over $\eta_i$, and making use of the symmetries in $t_2$ and $\tau$, one is left with
\begin{align}
S_{23} &= - S_{33}^\text{exc} + \nu \frac{4 e^2\Gamma_L^2\Gamma_R^2}{(2\pi a)^{2\nu + 2/
    \nu}} \int    d\tau  \cos (e^* V \tau) e^{2\nu\mathcal{G}(-\tau)}  \nonumber \\
    & \qquad \qquad \qquad  \times \int dt_2   \left( e^{\frac{2}{\nu}\mathcal{G} (t_{2})}  - e^{\frac{2}{\nu}\mathcal{G} (-t_{2})}
\right) \left\{   F_{+1} (t_2, \tau)  -  i \theta \tau \left[  F_0 (t_2, \tau) -  F_{+1} (t_2, \tau) \right] \right\} ,  \\
      S_{33}^\text{exc} &= -\frac{4 e^2 \Gamma_L^2\Gamma_R^2}{(2\pi a)^{2\nu + 2/
    \nu}} \int  d\tau \cos (e^* V \tau) e^{2\nu\mathcal{G}(-\tau)} \nonumber \\
     & \qquad \times \int dt_2 \left[ \left( e^{\frac{2}{\nu}\mathcal{G} (t_{2})} + e^{\frac{2}{\nu}\mathcal{G} (-t_{2})}
\right) - 2 i \theta t_2 \left( e^{\frac{2}{\nu}\mathcal{G} (t_{2})} - e^{\frac{2}{\nu}\mathcal{G} (-t_{2})}
\right)
    \right] \left[ F_0 (t_2, \tau) -  F_{+1} (t_2, \tau) \right] ,
\end{align}
where we introduced the shorter notation $\mathcal{G}( t) = \mathcal{G}^{-+} (0, t)$ along with $F_\frac{\eta_3-\eta_4}{2} (t_2,\tau) = F_{\eta_3, \eta_4} (t_2,\tau)$.

Carrying out the integration over $t_2$, one can show, after rather lengthy but straightforward derivations, that
\begin{align}
 \label{eq:cross1int}
S_{23} &= - S_{33}^\text{exc} + 
\nu \frac{8 \pi}{(\pi \theta)^2 } \frac{e^2\Gamma_R^2 \Gamma_L^2}{\left( 2 \pi a \right)^{2\nu+2/\nu}}
\int d\tau \cos \left( e^* V \tau \right)  e^{2 \nu \mathcal{G} (-\tau)} \left[  i H_0 (\tau) - \left( 1 + i \theta \tau \right) M_0 (\tau)  \right]   , \\
S_{33}^\text{exc} &= -
\frac{8 \pi}{(\pi \theta)^2 } \frac{e^2\Gamma_R^2 \Gamma_L^2}{\left( 2 \pi a \right)^{2\nu+2/\nu}} 
\int d\tau \cos \left( e^* V \tau \right)  e^{2 \nu \mathcal{G} (-\tau)} \left( 1+2 i \theta \tau \right) M_0 (\tau)   ,
 \label{eq:auto1int}
\end{align}
where
\begin{align}
M_0 (\tau) &= \text{Re} \left\{ \alpha^{2/\nu} 2 \pi (-1)^{1/\nu} \frac{\sinh^2 \left( \pi \theta \tau \right)}{\left[ \sinh \left( \pi \theta \tau + 3 i \alpha  \right)\right]^{2/\nu}} \right\} , \\ 
H_0 (\tau) &= (2 \alpha)^{2/\nu} e^{-\pi \theta \tau}  \sinh \left( \pi \theta \tau \right) \frac{\Gamma \left( 1+\frac{1}{\nu} \right) \Gamma \left( \frac{1}{\nu}\right)}{\Gamma \left( 1+\frac{2}{\nu}\right)}  \null_2 F_1 \left( 1, \frac{1}{\nu}, \frac{2}{\nu}, 1 - e^{-2\pi \theta \tau} \right)  .
\label{eq:defH0}
\end{align}

\subsection{Final $\tau-$integral}

We are now left with a set of three integrals remaining. Two of those can be carried out for generic values of the filling factor. Indeed, using known results,~\cite{gradshteyn14} one can readily compute the intermediate function
\begin{align}
    \kappa_0 (z) &= \int dv e^{i z v} \left( \frac{\sinh (i \alpha)}{\sinh (v+ i \alpha)}\right)^{2 \nu} \frac{1}{\left[ \sinh (v+3 i \alpha )\right]^{2/\nu-2}} \nonumber \\
    &=- 2 ^{2/\nu+2\nu-3} \alpha^{2\nu} (-1)^{1/\nu} e^{-\pi z/2} \frac{\left| \Gamma \left( \frac{1}{\nu} + \nu -1 +i \frac{z}{2}\right)\right|^2}{\Gamma \left( \frac{2}{\nu} +2 \nu -2 \right)}   ,
\end{align}
and from this obtain
\begin{align}
    \int d\tau \cos \left( e^* V \tau \right)  e^{2 \nu \mathcal{G} (-\tau)}  M_0 (\tau) &= \frac{(-1)^{1/\nu} \alpha^{2/\nu}}{\theta} \left[ \kappa_0 \left( \frac{e^* V}{\pi \theta}\right) +  \kappa_0 \left( - \frac{e^* V}{\pi \theta}\right)  \right]   ,
\end{align}
as well as
\begin{align}
    \int d\tau \cos \left( e^* V \tau \right)  e^{2 \nu \mathcal{G} (-\tau)} \theta \tau M_0 (\tau)  &=   \frac{(-1)^{1/\nu} \alpha^{2/\nu}}{\pi \theta}  \left[ \frac{1}{i} \left. \partial_z \kappa_0 (z)\right|_{z=\frac{e^* V}{\pi \theta}}  +\frac{1}{i} \left. \partial_z \kappa_0 (z)\right|_{z=-\frac{e^* V}{\pi \theta}}  \right]   .
\end{align}

The final integral remaining involves the function $H_0 (\tau)$ defined in Eq.~\eqref{eq:defH0}. While it cannot be performed formally, one can make progress by noticing that the hypergeometric function only involves integer coefficients thus enabling a finite order expansion. The resulting coefficients, however, depend on the inverse filling factor $1/\nu$ in a nontrivial way. While the expansion can be carried out for any Laughlin filling factor, we choose to show only the results at $\nu = 1/3$. One then has
\begin{align}
    \null_2 F_1 \left( 1, 3, 6, 1 - e^{-2\pi \theta \tau} \right) &= \frac{- 30}{12\left( e^{-2\pi \theta \tau} -1\right)} \left[ \frac{24 \pi \theta \tau e^{-4 \pi \theta \tau}}{\left( e^{-2\pi \theta \tau} -1 \right)^4}  +  \frac{12}{\left( e^{-2\pi \theta \tau} -1 \right)^3} + \frac{18}{\left( e^{-2\pi \theta \tau} -1 \right)^2} + \frac{4}{\left( e^{-2\pi \theta \tau} -1 \right)} - 1 \right]  .
\end{align}
Substituting this back into Eq.~\eqref{eq:defH0}, one is able to carry out the only remaining integral, namely
\begin{align}
    \int d\tau \cos \left( e^* V \tau \right)  e^{\frac{2}{3}  \mathcal{G} (-\tau)}  H_0 (\tau) = i \frac{(2 \alpha)^{20/3}}{8 \pi \theta} & \left(  
\frac{\left| \Gamma\left(\frac{7}{3}+ i \frac{eV}{6\pi\theta} \right)\right|^2}
        {\Gamma\left(\frac{14}{3}\right)} 
        \left\{
\pi \cosh \left(\frac{eV}{6\theta}\right)
        -2 \sinh\left(\frac{eV}{6\theta}\right)  \text{Im}\left[\psi\left(\frac{7}{3} + i\frac{eV}{6\pi\theta}\right) \right]
        \right\} \right. \nonumber \\
        & \left.
- \frac{\left| \Gamma\left(\frac{1}{3}+ i \frac{eV}{6\pi\theta} \right)\right|^2}{\Gamma\left(\frac{5}{3}\right)} \frac{eV}{6\pi\theta}\sinh\left( \frac{eV}{6\theta}\right) \left[\frac{23}{120} + \frac{9}{160}\left( \frac{eV}{3\pi\theta}\right)^2  \right]
    \right)   .
\end{align}

Combining all these results, this finally leaves us with the final expressions for the current correlations at finite temperature and filling factor $\nu=1/3$
\begin{align}
        S_{23} &= 
        - \frac{2}{3} \frac{4 e^2\Gamma_L^2\Gamma_R^2}{\pi^2 v_F^3} \left( \frac{\theta}{v_F} \right)^{11/3} 
         \left\{ \sinh\left(\frac{eV}{6\theta}\right) \frac{\left| \Gamma\left(\frac{7}{3}+ i \frac{eV}{6\pi\theta} \right)\right|^2}
        {\Gamma\left(\frac{14}{3}\right)} \text{Im}\left[\psi\left(\frac{7}{3} + i\frac{eV}{6\pi\theta}\right) \right]   \right. \nonumber \\
        & \hspace{2cm} - 
        \left. \frac{1}{8} \frac{\left| \Gamma\left(\frac{1}{3}+ i \frac{eV}{6\pi\theta} \right)\right|^2}{\Gamma\left(\frac{5}{3}\right)} \frac{eV}{6\pi\theta}\sinh\left( \frac{eV}{6\theta}\right) \left[\frac{23}{120} + \frac{9}{160}\left( \frac{eV}{3\pi\theta}\right)^2  \right]\right\}, \\
        S_{33}^\text{exc} &= \frac{4 e^2\Gamma_L^2\Gamma_R^2}{\pi^2 v_F^3} \left( \frac{\theta}{v_F} \right)^{11/3}\sinh\left(\frac{eV}{6\theta}\right)\frac{\left| \Gamma\left(\frac{7}{3}+ i \frac{eV}{6\pi\theta} \right)\right|^2}{\Gamma\left(\frac{14}{3}\right)}\text{Im}\left[\psi\left(\frac{7}{3}+ i \frac{eV}{6\pi\theta} \right) \right]   .
\end{align}

\end{document}